\documentclass[aps,prl,nofootinbib,preprintnumbers,amsmath,amssymb,latexsym,array,enumerate,letter,twocolumn,superscriptaddress]{revtex4}
\usepackage{amssymb}
\usepackage{amsmath}
\usepackage{epsfig}
\usepackage{hyperref}
\usepackage{breakurl}
\usepackage{mathtools}
\usepackage{extpfeil}
\usepackage{comment}
\usepackage{graphicx,float,color}
\usepackage{ulem}
\usepackage{multirow}

\makeatletter
\def\simgt{\mathrel{\lower2.5pt\vbox{\lineskip=0pt\baselineskip=0pt
           \hbox{$>$}\hbox{$\sim$}}}}
\def\simlt{\mathrel{\lower2.5pt\vbox{\lineskip=0pt\baselineskip=0pt
           \hbox{$<$}\hbox{$\sim$}}}}
\makeatother

\newcommand{\be}{\begin{eqnarray}}
\newcommand{\ee}{\end{eqnarray}}
\newcommand{\bea}{\begin{eqnarray}}
\newcommand{\eea}{\end{eqnarray}}
\newcommand{\beq}{\begin{eqnarray}}
\newcommand{\eeq}{\end{eqnarray}}

\newcommand{\dd}[1]{\frac{\partial}{\partial #1}}

\def\lsim{\mathrel{\rlap{\lower4pt\hbox{\hskip1pt$\sim$}}
     \raise1pt\hbox{$<$}}}         
\def\gsim{\mathrel{\rlap{\lower4pt\hbox{\hskip1pt$\sim$}}
     \raise1pt\hbox{$>$}}}

\newcommand{\PA}{{\rm PA}}
\newcommand{\PX}{{\rm PX}}

\newcommand{\bF}{\boldsymbol{F}}
\newcommand{\bPhi}{\boldsymbol{\Phi}}
\newcommand{\bC}{\boldsymbol{C}}

\newcommand{\bPA}{\boldsymbol{\PA}}

\newcommand{\bM}{\boldsymbol{M}}

\newcommand{\bPsi}{\boldsymbol{\psi}}

\newcommand{\stkout}[1]{\ifmmode\text{\sout{\ensuremath{#1}}}\else\sout{#1}\fi}

\usepackage{newfile}
\allowdisplaybreaks

\begin{document}

\title{Pulsar Polarization Array Limits on Ultralight Axion-like Dark Matter}

\author{Xiao Xue}
\altaffiliation{These authors contributed equally to this work.}
\affiliation{Institut de Física d’Altes Energies (IFAE), The Barcelona Institute of Science and Technology, Campus UAB, 08193 Bellaterra (Barcelona), Spain}
\affiliation{II. Institute of Theoretical Physics, Universit\"{a}t Hamburg, 22761 Hamburg, Germany}
\affiliation{Deutsches Elektronen-Synchrotron DESY, Notkestr. 85, 22607 Hamburg, Germany}

\author{Shi Dai}
\altaffiliation{These authors contributed equally to this work.}
\affiliation{Australia Telescope National Facility, CSIRO Space and Astronomy, PO Box 76, Epping, NSW 1710, Australia}

\author{Hoang Nhan Luu}
\affiliation{Donostia International Physics Center, Basque Country UPV/EHU, San Sebastian, E-48080, Spain}

\author{Tao Liu}
\altaffiliation{Corresponding author: taoliu@ust.hk}
\affiliation{Department of Physics and Jockey Club Institute for Advanced Study, The Hong Kong University of Science and Technology, Hong Kong S.A.R., China}

\author{Jing Ren}
\altaffiliation{Corresponding author: renjing@ihep.ac.cn}
\affiliation{Institute of High Energy Physics, Chinese Academy of Sciences, Beijing 100049, China}
\affiliation{Center for High Energy Physics, Peking University, Beijing 100871, China}

\author{Jing Shu}
\affiliation{School of Physics and State Key Laboratory of Nuclear Physics and Technology, Peking University, Beijing 100871, China}
\affiliation{Center for High Energy Physics, Peking University, Beijing 100871, China}
\affiliation{Beijing Laser Acceleration Innovation Center, Huairou, Beijing, 101400, China}

\author{Yue Zhao}
\affiliation{Department of Physics and Jockey Club Institute for Advanced Study, The Hong Kong University of Science and Technology, Hong Kong S.A.R., China}

\author{Andrew Zic}
\affiliation{Australia Telescope National Facility, CSIRO Space and Astronomy, PO Box 76, Epping, NSW 1710, Australia}

\author{N. D. Ramesh Bhat}
\affiliation{International Centre for Radio Astronomy Research, Curtin University, Bentley, WA 6102, Australia}

\author{Zu-Cheng Chen}
\affiliation{Department of Physics and Synergetic Innovation Center for Quantum Effects and Applications, Hunan Normal University, Changsha, Hunan 410081, China}
\affiliation{Institute of Interdisciplinary Studies, Hunan Normal University, Changsha, Hunan 410081, China}

\author{Yi Feng}
\affiliation{Research Center for Astronomical Computing, Zhejiang Laboratory, 311121 Hangzhou, Zhejiang, China}

\author{George Hobbs}
\affiliation{Australia Telescope National Facility, CSIRO Space and Astronomy, PO Box 76, Epping, NSW 1710, Australia}

\author{Agastya Kapur}
\affiliation{Australia Telescope National Facility, CSIRO Space and Astronomy, PO Box 76, Epping, NSW 1710, Australia}

\author{Richard N. Manchester}
\affiliation{Australia Telescope National Facility, CSIRO Space and Astronomy, PO Box 76, Epping, NSW 1710, Australia}

\author{Rami Mandow}
\affiliation{Department of Physics and Astronomy and MQ Research Centre in Astronomy, Astrophysics and Astrophotonics, Macquarie University, NSW 2109, Australia}
\affiliation{Australia Telescope National Facility, CSIRO Space and Astronomy, PO Box 76, Epping, NSW 1710, Australia}

\author{Saurav Mishra}
\affiliation{Centre for Astrophysics and Supercomputing, Swinburne University of Technology, Hawthorn, VIC, 3122, Australia}

\author{Daniel J. Reardon}
\affiliation{Centre for Astrophysics and Supercomputing, Swinburne University of Technology, Hawthorn, VIC, 3122, Australia}
\affiliation{ARC Centre of Excellence for Gravitational Wave Astronomy}

\author{Christopher J. Russell}
\affiliation{CSIRO Scientific Computing, Australian Technology Park, Locked Bag 9013, Alexandria, NSW 1435, Australia}

\author{Ryan M. Shannon}
\affiliation{Centre for Astrophysics and Supercomputing, Swinburne University of Technology, Hawthorn, VIC, 3122, Australia}
\affiliation{ARC Centre of Excellence for Gravitational Wave Astronomy}

\author{Shuangqiang Wang}
\affiliation{Xinjiang Astronomical Observatory, Chinese Academy of Sciences, Urumqi, Xinjiang 830011, China}
\affiliation{Australia Telescope National Facility, CSIRO, Space and Astronomy, PO Box 76, Epping, NSW 1710, Australia}

\author{Lei Zhang}
\affiliation{National Astronomical Observatories, Chinese Academy of Sciences, A20 Datun Road, Chaoyang District, Beijing 100101, China}
\affiliation{Centre for Astrophysics and Supercomputing, Swinburne University of Technology, Hawthorn, VIC, 3122, Australia}

\author{Songbo Zhang}
\affiliation{Purple Mountain Observatory, Chinese Academy of Sciences, Nanjing 210023, China}
\affiliation{Australia Telescope National Facility, CSIRO Space and Astronomy, PO Box 76, Epping, NSW 1710, Australia}

\author{Xingjiang Zhu}
\affiliation{Department of Physics, Faculty of Arts and Sciences, Beijing Normal University, Zhuhai 519087, China}
\affiliation{Institute for Frontier in Astronomy and Astrophysics, Beijing Normal University, Beijing 102206, China}
\affiliation{Advanced Institute of Natural Sciences, Beijing Normal University, Zhuhai 519087, China}

\collaboration{PPTA Collaboration}

\begin{abstract}

We conduct the first-ever pulsar polarization array (PPA) analysis to detect the ultralight Axion-Like Dark Matter (ALDM) using the polarization data of 22 millisecond pulsars from the third data release of Parkes pulsar timing array. As one of the major dark matter candidates, the ultralight ALDM exhibits a pronounced wave nature on astronomical scales and offers a promising solution to small-scale structure issues within local galaxies. While the linearly polarized pulsar light travels through the ALDM galactic halo, its position angle (PA) can be subject to an oscillation induced by the ALDM Chern-Simons coupling with electromagnetic field. The PPA is thus especially suited for detecting the ultralight ALDM by correlating polarization data across the arrayed pulsars. To accomplish this task, we develop an advanced Bayesian analysis framework that allows us to  construct pulsar PA residual time series, model noise contributions properly and search for pulsar cross-correlations. We find that for an ALDM density of $\rho_0=0.4\,\textrm{GeV}/\textrm{cm}^3$, the Parkes PPA offers the best global limits on the ALDM Chern-Simons coupling, namely $\lesssim 10^{-13.5}-10^{-12.2}~{\rm GeV}^{-1}$, for the mass range of $10^{-22} - 10^{-21}~{\rm eV}$. The crucial role of pulsar cross-correlation in recognizing the nature of the derived limits is also highlighted. 

\end{abstract}
\preprint{DESY-24-100}
\maketitle

\section{Introduction}

Millisecond pulsars (MSPs) are known as astronomical clocks due to the remarkable regularity of their pulses. The Pulsar timing array (PTA), consisting of many accurately timed MSPs in the Milky Way, thus has been proposed as a galactic-scale detector for nanohertz gravitational waves (GWs)~\cite{detweiler1979pulsar, foster1990constructing}. 
The Hellings-Downs (H-D) curve~\cite{hellings1983upper} plays a crucial role in recognizing the signals of stochastic GW background (SGWB), as it manifests the unique correlation of signal-induced timing residuals across the MSPs. Recently, four leading PTA collaborations~\cite{NANOGrav:2023gor,EPTA:2023fyk,Reardon:2023gzh,Xu:2023wog} reported a  consistency between their data and the H-D curve, strongly supporting the SGWB interpretation of the observed common process excess. This milestone for GW astronomy marks a roaring success for the PTA programs.

Pulsars are also known as excellent astronomical linear polarization sources. Their light can be linearly polarized to wellabove $10\%$ of total intensity. The PTA programs usually measure the pulsar polarization profiles for calibrating the relevant observations~\cite{oslowski2013improving}. Vast amounts of polarization data thus exist or are expected to be collected in the active PTAs and other pulsar-monitoring programs. This has motivated the suggestion of developing pulsar polarization arrays (PPAs)~\cite{Liu:2021zlt}, to explore astrophysics and fundamental physics with a correlated polarization signal across the pulsars. One representative scientific case is detecting ultralight axion-like dark matter (ALDM).

Dark matter (DM) makes up $\sim 85\%$ of the matter in the universe, yet its nature remains uncertain. The ALDM is one of the major theories to explain the DM observations. The concept of an axion arose originally to address the 
strong charge-parity problem in quantum chromodynamics, while many particle physics theories predict axion-like particles. The ultralight ALDM, due to its long 
de Broglie wavelength, 
is distinguished from the other DM candidates by its wave nature at astronomical scales. 
Of particular interest is the ultralight ALDM with a mass of $10^{-22}-10^{-21}\,$eV, generally known as ``fuzzy DM''~\cite{Hu:2000ke,hui2017ultralight}, which offers a potential solution to the ``small-scale'' structure problems~\cite{weinberg2015cold,Hu:2000ke,hui2017ultralight}. The constraints on this DM scenario may arise from the observations of Lyman-$\alpha$ forest and dwarf galaxies (for a review, see~\cite{Ferreira:2020fam}). As these constraints are under debate~\cite{Zhang:2017chj,Ferreira:2020fam,Hayashi:2021xxu,Dalal:2022rmp}, it is highly valuable to develop independent probes.

The PPA is especially suited for the detection of the ultralight ALDM. Due to condensation, the ALDM galactic halo can be described as a classical field~\cite{Hui:2021tkt}. The ALDM field couples to an electromagnetic (EM) field via the Chern-Simons term $\sim \frac{1}{2}g_{a\gamma\gamma}\,a\,F_{\mu\nu}\tilde{F}^{\mu\nu}\,$. Here $a$ is the ALDM field, $F_{\mu\nu}$ is the EM field strength and $\tilde{F}_{\mu\nu}$ is its Hodge dual, and $g_{a\gamma\gamma}$ is the Chern-Simons coupling. The ALDM halo as a background breaks the degeneracy of dispersion relation between the left- and right-circular polarization modes of light because of its odd parity. As a result, while the linearly polarized pulsar light travels through the ALDM halo, its position angle (PA) can rotate. This effect in a general context is known as ``cosmic birefringence (CB)''~\cite{Carroll:1989vb,Carroll:1991zs}.  The ALDM wave nature at astronomical scales predicts a signature of oscillating PA with specific pattern in spacetime~\cite{Liu:2021zlt}. The PPA by cross-correlating the data is thus expected to be highly capable of recognizing the ultralight ALDM signals.

In this letter, we conduct the first-ever PPA analysis to search for the ultralight ALDM, using the polarization data of 22 MSPs from PPTA Data Release~3 (DR3)~\cite{Zic:2023gta, Reardon:2023gzh}. We denote this PPA as the Parkes PPA (PPPA) for the convenience of discussions. To accomplish this task, we develop an advanced analysis framework  tailored specifically for constructing real pulsar PA residual time series,
modeling instrumental and astronomical noises and searching for signal correlation across the arrayed pulsars. Using this Bayesian framework, we demonstrate that the PPPA provides the most stringent constraints on the ALDM Chern-Simons coupling within the relevant mass range. Notably, we demonstrate for the first time that the pulsar cross-correlation plays a crucial role in investigating the nature of the derived limits.

\section{ALDM induced cosmic birefringence}

The DM halos originate from primordial density fluctuations, with the DM particles becoming virialized over the galaxy history~\cite{White:1978}. Consequently, the ultralight ALDM field can be viewed as a random superposition of a vast number of plane waves~\cite{Foster:2017hbq,Centers:2019dyn,Foster:2020fln,Nakatsuka:2022gaf}: 
\begin{eqnarray}\label{eq:axionf}
a(\mathbf{x},t) &\approx& \frac{\sqrt{\rho(\mathbf{x})}}{m_a}   \sum_{\mathbf{v}\in\Omega}(\Delta v)^{3/2}\alpha_{\mathbf{v}}\sqrt{f_{\mathbf{x}}(\mathbf{v})}\nonumber\\
&&\times \cos[m_a(t-\mathbf{v}\cdot\mathbf{x})+\phi_{\mathbf{v}}]\,, 
\end{eqnarray}
where $\rho(\mathbf{x})$ is ALDM energy density at position $\mathbf{x}$, $m_a$ is ALDM mass, $\mathbf{v}$ denotes velocity in a latticed phase space $\Omega$, and $\Delta v$ is lattice spacing. $f_{\mathbf{x}}(\mathbf{v})$ represents DM velocity distribution at $\mathbf{x}$, 
which is assumed to be spatially uniform and isotropic, with the speed $|\mathbf{v}|$ peaking at the halo's virial velocity, {\it i.e.} $v_0\sim \mathcal O(100)\,$km/s.
The profile of $\rho(\mathbf{x})$ is determined by Schr\"{o}dinger-Poisson equations. As shown in~\cite{Schive:2014dra}, the halo self-gravitation leads to the formation of a central solitonic core within galaxies. Since most PPPA MSPs are only $\lesssim \mathcal O(10^3)$ parsec away from the Earth, 
we simply assume $\rho(\mathbf{x})$ to be uniform, {\it i.e.}, $\rho(\mathbf{x})= \rho_0$~\cite{footnote1}, 
with its value to be determined by galactic halo measurements.
The parameters $\alpha_\mathbf{v} \in (0, +\infty)$ and $\phi_\mathbf{v} \in (0, 2\pi)$ represent random amplitude and phase, respectively, following Rayleigh and uniform distributions~\cite{Foster:2017hbq}. For a specific realization of $\{\alpha_\mathbf{v}, \phi_\mathbf{v}\}$, the ALDM field $a(\mathbf{x}, t)$ exhibits stochastic spacetime dependence~\cite{Centers:2019dyn}, where coherence gets lost for a temporal interval $\gg \tau_c$ or a spatial interval $\gg l_c$, with $\tau_c \sim 1/(m_av_0^2)$ and $l_c\sim 1/(m_a v_0)$ denoting the coherent time and coherent length (i.e. de Broglie wavelength) of this field. For the mass range of ultralight ALDM, 
the observation period $T_{\rm obs}$ for existing PTA programs is significantly shorter than $\tau_c$, leading to coherent temporal evolution of $\cos(m_a t + \ldots)$ within this period as shown in Eq.~(\ref{eq:axionf}). However, the distances of pulsars, whether to another pulsar or to the Earth, may exceed $l_c$, resulting in stochastic variation of the ALDM field over such distances.

Given the ALDM field within the Milky Way, one can calculate the induced PA residual $\Delta\PA^a$ for the linearly polarized pulsar light. As the CB effect is topological~\cite{Harari:1992ea}, $\Delta\PA^a$ depends on the $a(\mathbf{x},t)$ values solely at two endpoints of the light path. For a pulse emitted at $(\mathbf{x}_p, t_p)$ from the pulsar and received at $(\mathbf{x}_e, t_e)$ on the Earth, we have   
\begin{align}
    \Delta\PA^a \approx g_{a\gamma\gamma} \left[a(\mathbf{x}_p, t_p)-a(\mathbf{x}_e, t_e)\right]\, ,  \label{eq:CB}
\end{align}
where the two terms are dubbed ``pulsar'' and ``Earth'' terms. 
The $\Delta\PA^a$ magnitude can be characterized by  a parameter $S_a \equiv g_{a\gamma\gamma}\sqrt{\rho_0}/m_a$ then. For $m_a\sim 10^{-22}\,$eV and $\rho_0\sim 0.1\,\textrm{GeV}/\textrm{cm}^3$, this implies that a signal of $\Delta\PA^a \sim 0.1\,$deg needs the Chern-Simons coupling to be $g_{a\gamma\gamma}\sim 10^{-13}\,\textrm{GeV}^{-1}$. Note, while the SGWB-induced timing residual in the PTA is also determined by so-called ``pulsar'' and ``Earth'' terms, unlike the CB effect, it arises from a cancellation between contraction and expansion of the light path as the GWs pass by. Moreover, the CB is distinguished from the Faraday rotation in interstellar medium (ISM) and Earth's ionosphere as the latter effect is frequency dependent and relies on the length of light path aligned with the direction of magnetic field.

\section{PPA analysis framework.}
\label{sec:3}

In last decade, a few attempts have been made to explore temporal variations of pulsar polarization~\cite{2010ApJ...721.1044W, Yan:2011bq,NANOGrav:2021yup}. The PA residuals of individual pulsars have been suggested for testing the ALDM theory~\cite{Liu:2019brz, Caputo:2019tms}. 
However, a systematic analysis framework of the PPA, which emphasizes the data correlation across the pulsars, remains absent. To address this, we model the 
PA residuals as: 
\begin{align}\label{eq:PAmodel}
\Delta\PA^{\rm obs} = 
    \Delta \PA^a + \Delta\PA^{\rm ion} + \Delta \PA^w + \Delta \PA^r + \Delta\PA^{\rm pol}\, .
\end{align}
Besides the stochastic signal  $\Delta {\rm PA}^a$ defined in Eq.~(\ref{eq:CB}), there are additional components categorized as deterministic and random. The deterministic components include the Faraday rotation induced by the Earth's ionosphere, $\Delta {\rm PA}^{\rm ion}$, which varies over timescales from hours to years and is derived with the package \texttt{ionFR}. 
The random contributions consist of white noise, $\Delta {\rm PA}^{w}$, which incorporates radiometer noise and accounts for random pulse-to-pulse variations in polarization profiles, along with unmodeled ionospheric effects that persist after the \texttt{ionFR} subtraction on shorter timescales.
Red noise, $\Delta {\rm PA}^{r}$, is also random and captures variations in ISM Faraday rotation or fluctuations in dispersion measure and rotation measure and unmodeled ionospheric effects over years using a frequency-domain power spectrum.
Additionally, we model long-term effects observed in the data with a polynomial of up to second order, denoted by $\Delta {\rm PA}^{\rm pol}$, which may account for slow stochastic changes in ISM properties over decades. (see Supplemental Material (SM) at Sec.~B for details).


The ALDM signal nature can be encoded in its covariance matrix $\bC^a$, as a two-point correlation function of $\Delta \PA^a$~\cite{Liu:2021zlt}. In the data, $\Delta \PA^a$ is constructed as a vector $\{\Delta\PA_{p,n}^a\}$, where $p$ runs over all PPA pulsars and $n$ runs over all epochs for each of them. Inheriting from the ALDM properties~\cite{Foster:2020fln}, this vector follows a multivariate Gaussian distribution with zero mean. By integrating the velocity of ALDM particles with a delta-function approximation for the speed at $|\mathbf{v}|=v_0$, 
we find $C_{p,n;q,m}^a \approx S^2_a\, \hat{C}_{p,n;q,m}^a\,$~\cite{Liu:2021zlt},where
\begin{eqnarray}\label{Eq:axion_cov}
\hat{C}_{p,n;q,m}^a &=& 
\cos[m_a(t_{p,n} - t_{q,m})] \nonumber\\
&&
+\cos[m_a(t'_{p,n}  - t'_{q,m})]{\rm sinc}(y_{pq})\nonumber\\
&&-\cos[m_a(t'_{p,n} - t_{q,m})]{\rm sinc}(y_{ep})\nonumber\\&&
-\cos[m_a(t_{p,n}  - t'_{q,m})]{\rm sinc}(y_{eq}) 
\end{eqnarray}
reflects quadratic interplay of its Earth and pulsar terms (see Eq.~(\ref{eq:CB})).  $\hat{C}_{p,n;q,m}^a$ is essentially a correlation measure. In its expression, $t_{p,n}$ and $t'_{p,n}=t_{p,n}-D_p$ denote the light reception and emission time for the $n$-th epoch of the $p$-th pulsar, with $D_{p}$ being the pulsar distance to the Earth, and their cosine functions denote temporal correlations of pulsar signals. 
The sinc functions ${\rm sinc}(y_{ij}) \equiv \sin(y_{ij})/y_{ij}$ are all related to the pulsar term and manifest signal spatial correlations. Here, $y_{ij} \equiv |\mathbf{x}_{i}-\mathbf{x}_{j}|/l_c$ is a dimensionless distance parameter.
These spatial correlations get suppressed when $|\mathbf{x}_{i}-\mathbf{x}_{j}|$ is significantly greater than the coherence length $l_c$.

These correlations originate from the spacetime dependence of the ALDM field phase factor (see Eq.~(\ref{eq:axionf})) or equivalently the ALDM wave nature. All terms in $\hat{C}_{p,n;q,m}^a$ can contribute significantly to recognizing the ALDM signals~\cite{Liu:2021zlt}. Due to its nonrelativistic nature, the ALDM de Broglie wavelength 
is approximately three orders of magnitude longer than its Compton wavelength ($\sim 1/m_a$). 
The nanoHz ($\sim 10^{-23.4}\,$eV) ALDM thus has a de Broglie wavelength  $\sim 1.5\,$kpc, yielding $y_{ij} \lesssim 1$
for many PPPA pulsars. This outcome contrasts with the PTA response to the nanoHz SGWB, where the H-D curve arises from the interplay between the Earth terms, and the pulsar terms are less important since the de Broglie wavelength of GWs is simply their Compton wavelength and hence much shorter than general pulsar distances for the current PTAs.

For the PPA analysis, the matrix $\bC^a$ defines a signal model with full correlation, with its diagonal ($p=q$) and off-diagonal blocks ($p\neq q$) manifesting the pulsar autocorrelation and cross-correlation, respectively. Below, we will focus on the full-correlation signal model, and as a reference, consider also the autocorrelation-only model where $\hat{C}^{a,{\rm auto}}_{p,n;q,m} = \hat{C}^{a}_{p,n;q,m}\delta_{pq}$.

The sensitivity analysis is based on Bayesian statistics. 
It is performed using the likelihood function~\cite{Liu:2021zlt}
\begin{align}
     \ln\mathcal{L} =
    -\frac{1}{2}\widetilde{\Delta\PA}^T \bC^{-1}\widetilde{\Delta\PA}- \frac{1}{2}\ln |2\pi\bC|\, .
    \label{Eq:likelihood}
\end{align}
Here $\widetilde{\Delta\PA}\equiv\Delta\PA^{\rm obs}  - \Delta\PA^{\rm ion}-\Delta\PA^{\rm pol}$ is the time series of PA residuals with $\Delta\PA^{\rm ion}$ and $\Delta\PA^{\rm pol}$ subtracted. 
$\bC=C^w+C^r+S_a^2 \hat{C}^a$ is the full covariance matrix that includes the contributions of the white and red noises and the ALDM signals. 
For the parameter $S_a$, we set the average ALDM energy density $\rho_0=0.4\,\textrm{GeV}/\textrm{cm}^3$~\cite{2010A&A...523A..83S} as the benchmark, since the Milky Way inner region ($\sim 20\,$kpc), where its rotation curve is measured for determining the local DM density~\cite{Nesti:2013uwa}, is considerably larger than $l_c$ for the mass range considered below.
Then by integrating out the $\Delta\PA^{\rm pol}$ parameters we obtain a likelihood parameterized by $\{\theta_n, D_p^*, m_a, S_a\}$, where $\theta_n$ denotes the set of random noise parameters, and $D_p^*$ is a pulsar distance normalized by the measured central value. To determine the spatial correlations in Eq.~(\ref{Eq:axion_cov}), we account for the $D_p^*$ uncertainties using priors determined by the measurement methods. For any given ALDM mass, we then perform a Bayesian analysis to constrain $S_a$ with this marginalized likelihood. Eventually, we convert the posterior distributions of $S_a$ to the limits on the ALDM Chern-Simons coupling for the given benchmark value of $\rho_0$. In addition to imposing limits, we use two Bayes factors ${\rm BF}^{\rm Full}_{\rm Null}$ and ${\rm BF}^{\rm Full}_{\rm Auto}$ to compare the full-correlation signal model with the null-signal model and to differentiate it from the autocorrelation-only signal model, respectively. The latter is particularly crucial for understanding the nature of the derived limits and recognizing the real signals in the data (see SM at Secs.~A and B for details).

\begin{figure}[htb]
    \centering
    \includegraphics[scale=0.7]{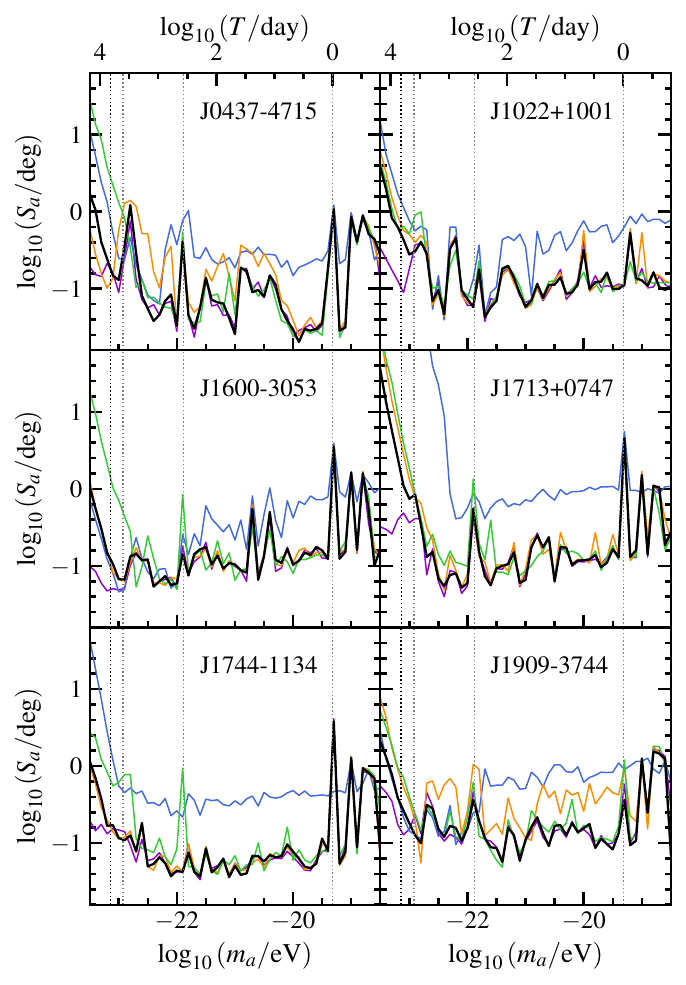}
    \caption{95\% upper limits on the ALDM characteristic signal strength $S_a$ as a function of its mass $m_a$, for six representative PPPA MSPs. In each panel, the black solid line represents the single-pulsar limits with full noise modeling in Eq.~(\ref{eq:PAmodel}), while the colored lines show the limits excluding individual noise components: additional white noise (blue), red noise (orange), linear and quadratic deterministic noise (purple), and ionospheric effects (green). The difference between the black and colored lines reflects the contribution of each noise component to the analysis.
    As a reference, we present  $T_{\rm obs}$ of each pulsar, and timescales of solar cycle (11 years), one year and one day, as colored and gray vertical dotted lines, respectively. The timescale and $m_a$ are related by $m_a = 2\pi/T$.}
    \label{fig:Salimit}
\end{figure}

\section{PPA results and discussions}

The PPTA DR3~\cite{Zic:2023gta} includes the measurements of Stokes parameters for 32 MSPs in the radio band, with a period from 2004 to 2022. We reprocess all data taken with the Ultra-Wideband Low receiver system, with the most up-to-date polarization calibration model to improve the measurement quality~\cite{2020PASA...37...12H}, and  
correct for static Faraday rotations caused by the ISM using the rotation measures provided by the Australia Telescope National Facility pulsar catalog~\cite{Manchester:2005}. We then select 22 MSPs for the PPPA construction whose signal-to-noise ratio of linear polarization intensity is relatively high in the considered frequency band ($2.6-3.6\,$GHz). For these MSPs, we align the observation in each epoch with the PPTA template of total intensity~\cite{Zic:2023gta,Dai:2015awa}. We then derive the standard PA profiles, subtract them from the data, and perform a weighted average over the phase bins, to generate PA residual time series for unbiased signal estimation (see SM at Sec.~A for details).

We then test the noise model for individual MSPs using the likelihood in Eq.~(\ref{Eq:likelihood}), with the ALDM signal turned off. To achieve this, we introduce a set of Bayes factors as a metric to assess the improvements in noise modeling by adding additional noise components one at a time to the minimal noise model, ultimately leading to the full noise model (see Table~II at  Sec.~B of SM for details).
Compared to the minimal model including only the observation error of $\Delta \PA^{\rm obs}$ and a constant deterministic residual~\cite{Liu:2021zlt}, the full noise model  shows a remarkable improvement in fitting the data for the majority of the 22 PPPA MSPs. 
The most significant improvements are observed for J0437-4715, J1713+0747 and J1909-3744. 
The contribution of each new component vary greatly among different pulsars. The analysis suggests that the imperfect modeling of ionospheric Faraday rotation may account for the additional white noise and even red noise observed for some specific pulsars, while slow change of the ISM properties over decades could explain the observed deterministic linear and quadratic variations. This outcome fully demonstrates the effectiveness of incorporating new noise components in modeling the observed PA residuals in Eq.~(\ref{eq:PAmodel}).

We demonstrate in Fig.~\ref{fig:Salimit} the 95\% upper limits on the ALDM characteristic signal strength $S_a$ as a function of $m_a$, for six representative PPPA MSPs individually. For the mass range of $m_a\gtrsim 1/T_{\rm obs}$, the $S_a$ limits are relatively flat. A best limit as strong as $0.1\,$deg or even stronger can be achieved for many of these pulsars including J0437-4715, J1744-1134, J1022+1001, etc. As the ALDM mass decreases to below $1/T_{\rm obs}$, the limits deteriorate approximately with $m_a^{-3}$, consistent with the expectation from integrating $\Delta\PA^{\rm pol}$ as a second-order polynomial expansion~\cite{Hazboun:2019vhv}.
The ionospheric subtraction effectively removes sharp peaks around $\sim 1 \,$yr or/and low-frequency excess for J0437-4715, J1600-3053 and J1744-1134. The incorporation of additional white noise component enhances the limits for six pulsars uniformly, and of red noise component improves intermediate- and low-frequency limits for J0437-4715 and J1909-3744 notably. These analyses show that, due to the variance of noise nature over pulsars, the advanced noise model in Eq.~(\ref{eq:PAmodel}) is essential for getting robust limits on the ALDM signal strength (see SM at Sec.~B for details).

\begin{figure}
    \centering
    \includegraphics[scale=0.5]{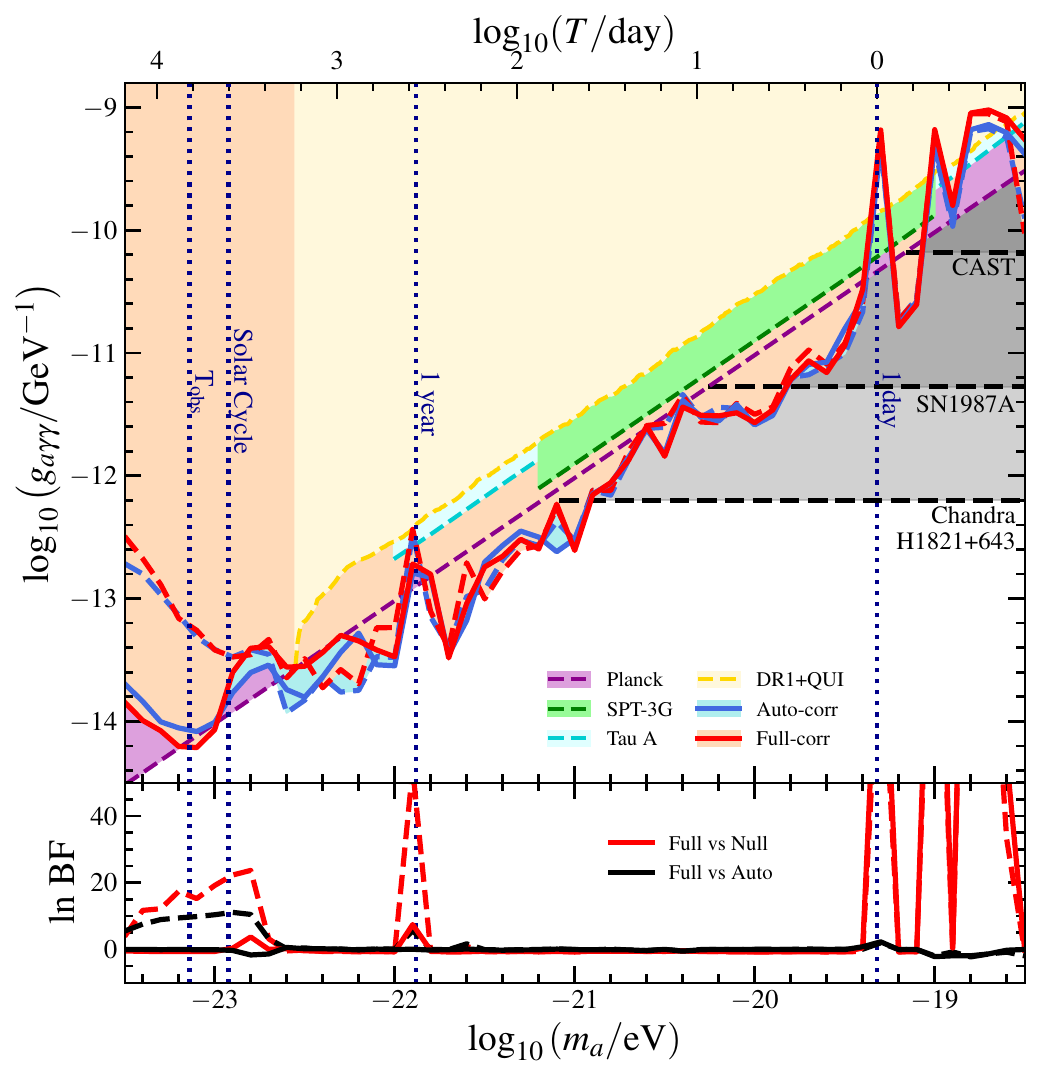}
    \caption{95\% upper limits of the PPPA on the ALDM Chern-Simons coupling $g_{a\gamma\gamma}$ as a function of its mass $m_a$, with $\rho_0=0.4\,\textrm{GeV}/\textrm{cm}^3$. The blue and red curves represent the limits obtained with the autocorrelation-only and full-correlation signal models, respectively, with (solid) and without (dashed) the ionospheric subtraction. As a reference, the smoothened constraints from the observation of PA variation in Crab Nebula with POLARBEAR~\cite{POLARBEAR:2024vel}, the analysis of PPTA data of DR1 and Crab Nebula data from QUIJOTE~\cite{Castillo:2022zfl}, the measurement of the CMB polarization with SPT-3G~\cite{SPT-3G:2022ods}, and the investigation of PLANCK mission on the ALDM-induced washout effects during recombination~\cite{Fedderke:2019ajk}  
    are presented as cyan, yellow, green, and purple dashed lines. The limits from the CAST experiment~\cite{CAST:2017uph} and the observations of  SN1987A~\cite{Payez:2014xsa} and x-ray spectral distortion in quasar H1821+643~\cite{Reynes:2021bpe} are also presented as black dashed lines. The shaded regions above these lines indicate the range of $m_a$ and $g_{a\gamma\gamma}$ that they have excluded. The vertical dotted lines denote the reference timescales, including the PPPA observation period $T_{\rm obs}\approx 18~{\rm yrs}$, solar cycle, one year and one day.  
    In the lower subpanel, the Bayes factors are displayed in two scenarios: the full-correlation signal model against the null-signal model (red) and the full-correlation signal model against the autocorrelation-only signal model (black). }
    \label{fig:results}
\end{figure}

The final PPPA limits on the ALDM Chern-Simons coupling $g_{a\gamma\gamma}$ are presented in Fig.~\ref{fig:results} as a function of its mass $m_a$, for both full-correlation and autocorrelation-only signal models. These limits exhibit a linear dependence on $m_a$ for $m_a\gtrsim 1/T_{\rm obs}$, while scale as $m_a^{-2}$ for $m_a\lesssim 1/T_{\rm obs}$. This feature should be expected from the scaling behavior of the $S_a$ limits depicted in Fig.~\ref{fig:Salimit}. Moreover, one can see that the ionospheric subtraction softens the peak at the timescale of 1\,year in these limits. These limits, however, start to oscillate strongly as the timescale decreases to one day or shorter, resulting in some sharp peaks for $m_a \gtrsim 10^{-19.4}\,$eV.

To recognize the nature of these derived limits, we also present two Bayes factors in this figure. The overall uniformity of $\ln {\rm BF}_{\rm Auto}^{\rm Full} \sim 0$ for the considered mass range implies that no significant ALDM signals have been detected in the PPPA data. Particularly, despite the peaks of $\ln {\rm BF}_{\rm Null}^{\rm Full}$ at the timescales of one day or shorter, the approximate flatness in the $\ln {\rm BF}_{\rm Auto}^{\rm Full}$ curve indicates a deficiency in the data of the ALDM-induced cross correlation. These sharp peaks in the limits are thus more likely to be caused by imperfect noise modeling. This discussion can be applied to the peak observed for the limits at $m_a \sim 10^{-22}\,$eV also. It 
highlights the crucial role played by the cross correlation in recognizing the nature of the data.

In Fig.~\ref{fig:results}, we also present two categories of additional limits as a reference. The first category does not rely on the ALDM assumption, including the limits from CAST~\cite{CAST:2017uph}, SN1987A~\cite{Payez:2014xsa} and H1821+643~\cite{Reynes:2021bpe}. Since the signals arise from relativistic axion-like particles, the generated limits are flat up to the energy scale of their production. The second category is to search for the ALDM-induced oscillations in polarization data, including the analyses of PPTA-QUIJOTE~\cite{Castillo:2022zfl}, CMB~\cite{SPT-3G:2022ods}, and Crab Nebula~\cite{POLARBEAR:2024vel}. The method in Ref.~\cite{Fedderke:2019ajk} is based on washout effects caused by ALDM-induced oscillations on the CMB during the recombination era, which benefits from the relatively high DM density in the early universe. As these studies are ALDM-based, the derived limits on $g_{a\gamma\gamma}$ exhibit similar scaling features with $m_a$ to that of the PPPA limits. 
The PPPA provides the most stringent limits across a wide mass range within this category, and varying the value of $\rho_0$ does not change their relative positions.
For a local density value of $\rho_0=0.4\,\textrm{GeV}/\textrm{cm}^3$, especially, the PPPA offers the best global limits, namely $\lesssim 10^{-13.5}-10^{-12.2}~{\rm GeV}^{-1}$, for the mass range of fuzzy DM of $10^{-22} - 10^{-21}~{\rm eV}$. 
Thanks to the 18-year observation time span of PPTA DR3, its excellent sensitivities extend to $\sim 10^{-23}\,$eV.

\section{Conclusion and outlook}

In this Letter, we presented the first-ever PPA analysis to detect the ultralight ALDM, using the polarization data from the PPTA DR3. 
We developed a systematic analysis framework for this task, including proper data construction, advanced noise modeling and robust Bayesian sensitivity analysis. The derived limits are found to be superior to the existing ones for the mass range of fuzzy DM and hence represents the globally best. Particularly, the data correlation across the pulsars plays a pivotal role in recognizing the nature of these limits. This study represents an initial effort to investigate the great physical potential of current and forthcoming polarization data of pulsars. Besides further improvements of the developed PPA analysis framework, especially its noise modeling, several significant scientific tasks can be readily recognized for future explorations.

First of all, we can further advance the PPA applications for the ALDM detection. 
The PPA as an innovative tool is universal and can be built up with polarization data from other PTA programs also and even with broad datasets of normal pulsars. 
Moreover, we can synergize the PPA and the PTA. The ALDM halo can periodically perturb the spacetime metric and hence be detected by the PTA~\cite{Khmelnitsky:2013lxt,DeMartino:2017qsa,Porayko:2018sfa,NANOGrav:2023hvm,EuropeanPulsarTimingArray:2023egv}. By constructing a PPA-PTA detector where the polarization and timing signals are correlated, the capability of a pulsar array could be further enhanced~\cite{Li:2025xlr}. 
Finally, the current PTA and PPA analyses could be extended from the radio band to high frequencies. With the pulsar data collected by Fermi Large Area Telescope, the first $\gamma$-ray PTA constraints on the SGWB have been delivered in~\cite{Fermi-LAT:2022wah} and on the ultralight ALDM in~\cite{Luu:2023rgg,Xia:2023hov}. 
We would expect the current and forthcoming x-ray missions, such as the Imaging X-ray Polarimetry Explorer~\cite{IXPE1} and the Enhanced X-ray Timing and Polarimetry~\cite{eXTP:2018anb}, to offer valuable insights on this task for the PPA. We leave these explorations to future work.  


\begin{acknowledgments}
\subsection*{Acknowledgments}
X.X. is supported by Deutsche Forschungsgemeinschaft under Germany’s Excellence Strategy EXC2121 “Quantum Universe” - 390833306. IFAE is partially funded by the CERCA program of the Generalitat de Catalunya. X.X. was partly funded by the grant CNS2023-143767. 
Grant CNS2023-143767 funded by MICIU/AEI/10.13039/501100011033 and by European Union NextGenerationEU/PRTR. T.L. is supported by the Collaborative Research Fund under Grant No. C6017-20G which is issued by Research Grants Council of Hong Kong S.A.R. J.R. is supported in part by the National Natural Science Foundation of China under Grant No.
12435005. J.S. is supported by Peking University under startup Grant No. 7101302974 and the National Natural Science Foundation of China under Grants No. 12025507, No.12150015; and is supported by the Key Research Program of Frontier Science of the Chinese Academy of Sciences (CAS) under Grants No. ZDBS-LY-7003. Y.Z. is supported by U.S. Department of Energy under Award No. DESC0009959. Z.C.C. is supported by the National Natural Science Foundation of China (Grant No. 12247176 and No. 12247112) and the innovative research group of Hunan Province under Grant No. 2024JJ1006. Parts of this work were undertaken with the support of the Australian Research Council Centre of Excellence for Gravitational Wave Discovery (CE230100016). L.Z. is supported by ACAMAR Postdoctoral Fellowship and the National Natural Science Foundation of China (Grant No. 12103069). XJZ is supported by the National Natural Science Foundation of China (Grant No.~12203004) and by the Fundamental Research Funds for the Central Universities.
\end{acknowledgments}


\section{Supplemental Materials}

The Supplementary Materials contain additional calculations and derivations in support of the results presented in this letter. Specifically, we discuss the data preparation for the PPA analysis in Sec.~A,  and then introduce the PPA analysis framework in Sec.~B that we have developed.

\section{A. Data Preparation}
The original data of PPTA DR3 contains the measurements of four Stokes parameters in pulsar fold-modes, spanning approximately from 2004 to 2022~\cite{2023PASA...40...49Z}. The data from Ultra-Wideband Low (UWL) receiver~\cite{2020PASA...37...12H} are accumulated since 2018. We reprocess the UWL data set using the most up-to-date polarization model following procedures described in \cite{2020PASA...37...12H}. Radio frequency interference (RFI) has been identified, masked, and flagged as part of the PPTA DR3 process in~\cite{2023PASA...40...49Z}, whose data products were utilized in this study. A two-stage RFI mitigation procedure was applied to the UWL data sets to reduce interference across the wide frequency band, while the data sets prior to UWL were reprocessed using more advanced RFI excision techniques~\cite{2023PASA...40...49Z}.\footnote{ The RFI is complex due to a mixture of diverse sources, including radar, satellites, and various local electronic facilities. While a mitigation procedure was applied to remove easily identifiable RFI from the data, we anticipate that more subtle RFI may still be present in the datasets, contributing to both temporally uncorrelated and correlated noise. We leave this investigation to a future work.}

The Stokes parameters are measured as an average over either 0.5 or $1$ hour at each epoch. After subtracting Faraday rotation caused by the ISM, we average the observation for each epoch over the considered frequency band. We finally have Stokes parameters as a function of the pulse phase and observational epoch. Note that, for the ALDM mass of interest, {\it i.e.}, $m_a\lesssim 10^{-18}\,$eV, the induced signal oscillates with a period longer than hours. The relevant information is thus expected to be well-preserved by the data processed in this procedure.

In this article, we focus our analysis on the frequency band at $10$cm, which has a central frequency of $3100$ MHz with a bandwidth of 1GHz. This frequency band is relatively high for the PPTA data, subjecting to relatively weak Faraday rotation. To keep this choice consistent, we use the data only within the same frequency range for the UWL dataset. Finally, we chose 22 MSPs out of 32 in PPTA DR3 to construct the PPPA with significant linear polarization SNR in this frequency band.  The observations of PSR~J1713+0747 during its recent event of pulse profile change~\cite{2021ATel14642....1X,2021MNRAS.507L..57S,2024ApJ...964..179J} have been excluded.
We display the basic information on these PPPA pulsars in Table~\ref{tab:pulsar_properties}. 

\begin{table*}
	\centering
	\begin{tabular}{|c|c|ccc|c|cc|cc|c|}
		\hline
        & \multirow{2}{*}{Pulsar} & \multicolumn{3}{c|}{Observations} & Ecliptic Latitude & \multicolumn{2}{c|}{Pulsar Distance} & med & SD& \multirow{2}{*}{Refs}\\
        \cline{3-10}
		& & $T_{\rm obs}$(year)& $N_{\rm obs}$ & $ {\rm max}(\tilde{L}) / \sigma_{\tilde{L}}$ & (deg) & $\widetilde {\rm DM}$\,(kpc)& PX\,(mas) &($\sigma_{n}/$deg) & ($\Delta\PA/$deg)& \\
		\hline
		1&J0437--4715&18.0&1122&976.4& -67.87 & 0.156&6.37(9)& 0.13 & 0.91 &\cite{jlh+93,rhc+16,ymw17}  \\
		2&J0613--0200&17.5&333&17.9& -25.41  & 1.024&1.01(9)&8.0 & 8.2&\cite{lnl+95,rsc+21,ymw17}  \\
        3&J0614--3329&2.8&37&42.8& -56.87  & 2.691&1.1(11)&4.5 & 3.2&\cite{rrc+11,cpb+23,ymw17} \\
		4&J0711--6830&17.8&371&67.2& -82.89  & 0.106&N/A&7.7 & 9.0&\cite{bjb+97,rsc+21,ymw17}  \\
        5&J1017--7156&11.4&348&112.0& -67.74  & 1.807&0.6(6)&3.5 & 4.1&\cite{kjb+12,rsc+21,ymw17}  \\
        \hline
		6&J1022+1001&17.9&334&579.0& -0.06  & 0.834&1.55(16)&0.86 & 1.2&\cite{cnst96,rsc+21,ymw17} \\
		7&J1024--0719&17.8&221&142.8& -16.04  & 0.382&0.93(5)&1.9 & 1.9&\cite{bjb+97,rsc+21,ymw17}  \\
		8&J1045--4509&16.7&274&77.0& -47.71   & 0.334&1.7(7)&2.5 & 2.6&\cite{bhl+94,rsc+21,ymw17}  \\
        9&J1125--6014&7.4&164&187.7& -55.66   & 0.988&0.6(33)&2.7 & 2.2&\cite{fsk+04,rsc+21,ymw17}  \\
        10&J1545--4550&10.1&185&438.7& -25.29   & 2.247&0.45(14)&0.92 & 1.1&\cite{bbb+13,rsc+21,ymw17}  \\
        \hline
		11&J1600--3053&18.0&305&355.9& -10.07  & 2.535&0.53(6)&0.98 & 1.3 &\cite{jbo+07,rsc+21,ymw17}  \\
		12&J1603--7202&18.0&293&110.7& -49.96  & 1.129&0.3(3)&4.7 & 5.4&\cite{llb+96,rsc+21,ymw17}  \\
		13&J1643--1224&17.9&274&128.1& 9.78  & 0.791&1.1(1)& 1.9 & 2.3&\cite{lnl+95,rsc+21,ymw17}  \\
		14&J1713+0747&17.1&319&1400.6& 30.70   & 0.919&0.763(21)&0.43 & 1.2&\cite{fwc93,rsc+21,ymw17} \\
		15&J1730--2304&17.9&226&530.9& 0.19   & 0.512&2.0(1)&0.86 & 1.4&\cite{lnl+95,rsc+21,ymw17}  \\
    \hline
		16&J1744--1134&18.0&329&1206.9& 11.81  & 0.148&2.44(5)&0.39 & 0.72&\cite{bjb+97,rsc+21,ymw17}  \\
		17&J1824--2452A&15.8&120&241.7& -1.55  & 3.702&0.200(11)&1.1 & 1.1&\cite{lbm+87,rsc+21,ymw17}  \\
		18&J1857+0943&17.6&182&71.8& 32.32    & 0.767&0.85(16)& 6.0 & 6.5&\cite{srs+86,rsc+21,ymw17} \\
		19&J1909--3744&18.0&918&2078.6& -15.16  & 0.564&0.86(1)&0.71 & 1.8&\cite{jbv+03,rsc+21,ymw17}  \\
		20&J1939+2134&16.3&162&77.4& 42.30    & 2.896&0.35(3)&2.0 & 1.9&\cite{bkh+82,rsc+21,ymw17} \\  
      \hline
		21&J2145--0750&17.8&299&140.6& 5.31   & 0.693&1.40(8)& 2.9 & 3.1&\cite{bhl+94,rsc+21,ymw17}  \\
        22&J2241--5236&12.0&355&119.9& -40.39    & 0.963&0.96(4)&5.2 & 4.6 &\cite{kjr+11,rsc+21,ymw17}  \\
		\hline
        \hline
	\end{tabular}
	\caption{Basic information on the 22 PPPA MSPs. $T_{\rm obs}$ and $N_{\rm obs}$ denote observation period and epoch number for each pulsar. The $\widetilde {\rm DM}$ column shows pulsar distances inferred from the dispersion measure. In the PX column, the numbers in brackets denote the last-digit errors of parallax measurements. The ``med'' column shows the median measurement error $\sigma_{n}$ of the observed PA residuals in Eq.~(\ref{eq:PAresObs}) for a given pulsar, and the ``SD'' column displays the standard deviation (SD) of the observed PA residuals after subtracting ionosphere corrections using the package \texttt{ionFR}~\cite{Sotomayor:2013} (see the ``Noise modeling'' part in Sec.~B for more details).
 }
 \label{tab:pulsar_properties}
\end{table*}

The total intensity $I$ exhibits the highest SNR among the Stokes parameters. 
So we utilize its standard profile $\tilde{I}$ provided by the PPTA~\cite{Dai:2015awa,2023PASA...40...49Z} and the package {\tt psrchive}~\cite{hotan2004psrchive} 
to align the profiles of all epochs.  
As the epochs are binned differently w.r.t. the phase in the original data, we rebin these profiles uniformly into 256 bins. To ensure high-quality data, we exclude the profiles where the phase offset exceeds $1/512$. Then for the  linearly polarized light in the $k$-th phase bin of the $n$-th observational epoch we have the intensity 
\begin{align}
    L_{n,k} \equiv \sqrt{(Q_{n,k})^2+(U_{n,k})^2}  \, ,
\end{align}
and the PA
\begin{align}
    \PA_{n,k}\equiv \frac{1}{2}\arctan\left(\frac{U_{n,k}}{Q_{n,k}}\right)\, ,
\end{align}
where  $\{I_{n,k},Q_{n,k},U_{n,k},V_{n,k}\}$ denote the four Stokes parameters. 
In this study, the noises of $Q_{n,k}$ and $U_{n,k}$ are assumed to be statistically independent and have the same errors $\sigma_{S,n}$ determined by the off-pulse RMS for each epoch. 

\subsection*{Standard profile of the linearly polarized pulsar light}

\begin{figure*}[ht]
    \centering
\includegraphics[width=0.4\textwidth]{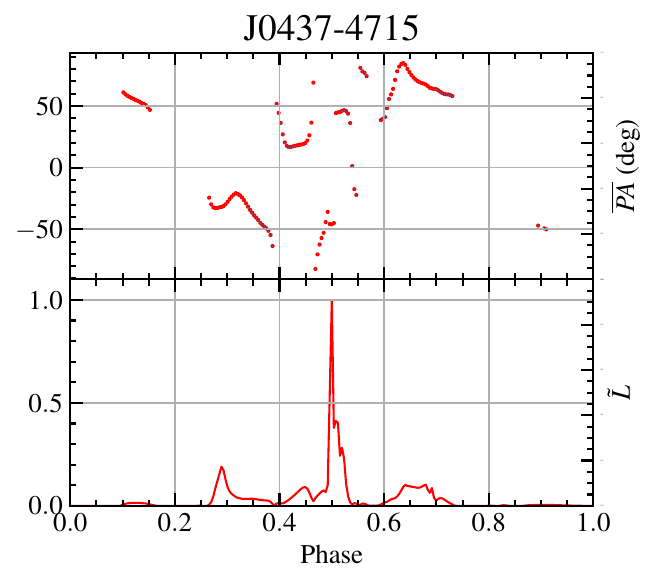}
\includegraphics[width=0.4\textwidth]{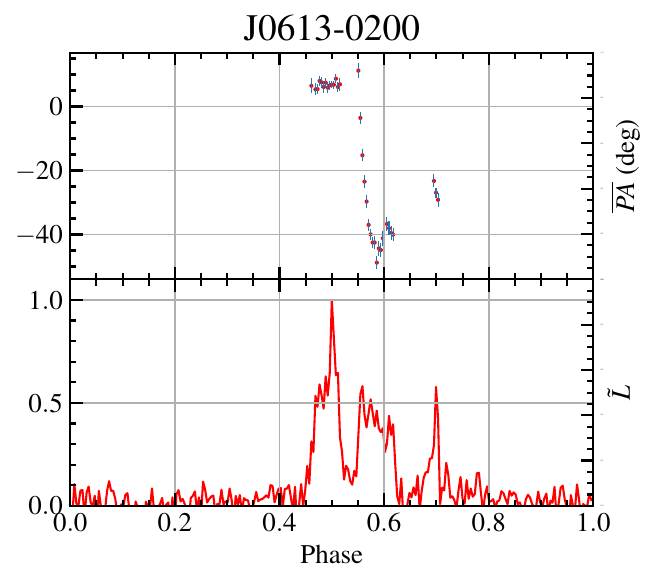}
\includegraphics[width=0.4\textwidth]{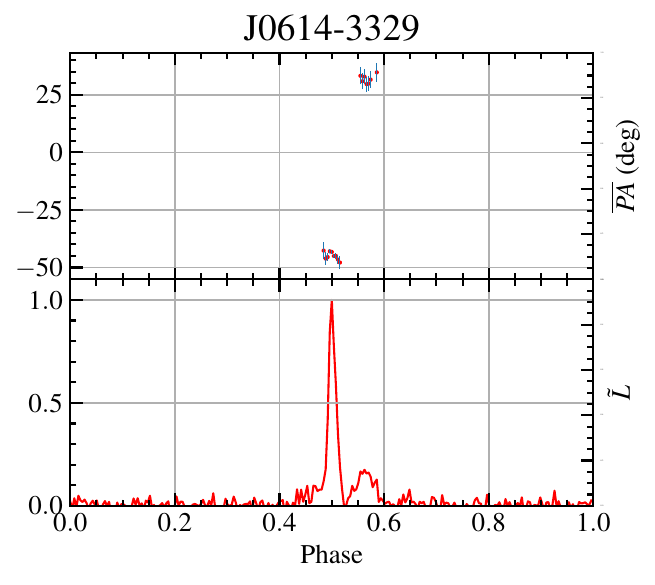}
\includegraphics[width=0.4\textwidth]{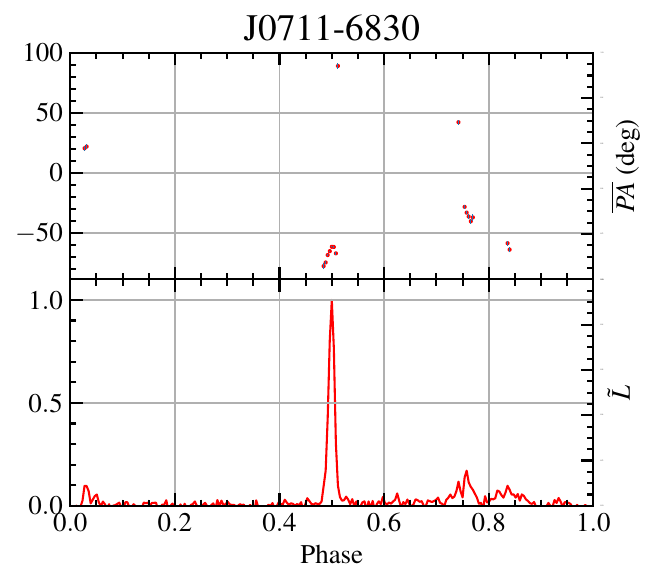}
\includegraphics[width=0.4\textwidth]{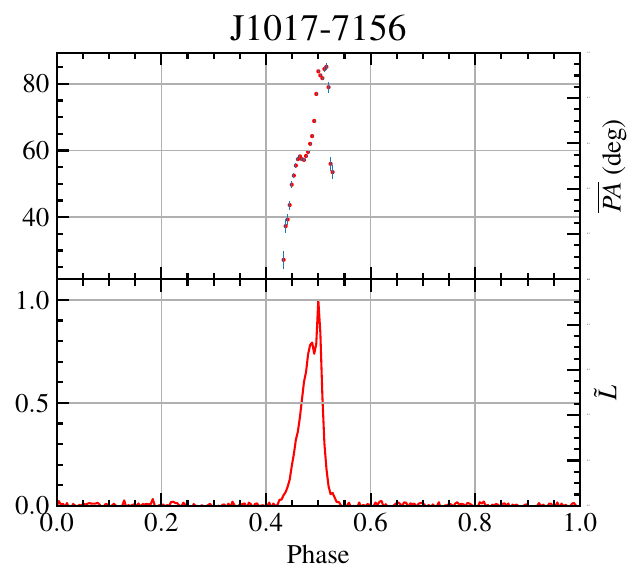}
\includegraphics[width=0.4\textwidth]{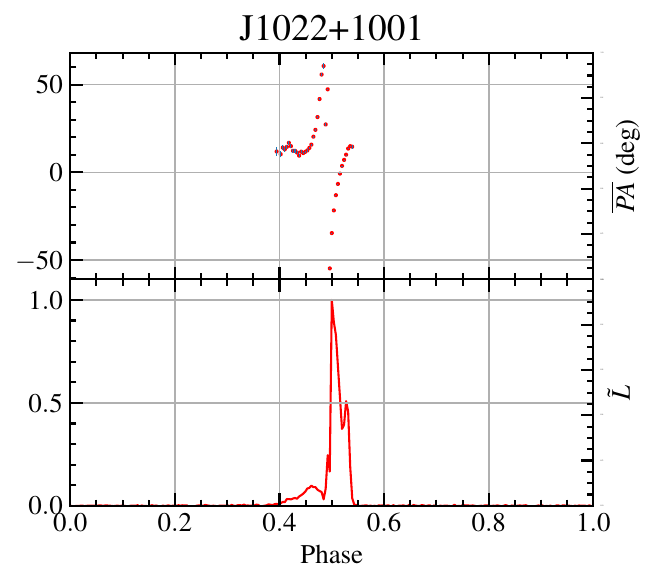}
\caption{Standard profiles of PA (upper sub-panel) and intensity (bottom sub-panel) for the linearly polarized light of PPPA MSPs. }
\label{fig:Profiles1}
\end{figure*}

\begin{figure*}[ht]
    \centering
\includegraphics[width=0.4\textwidth]{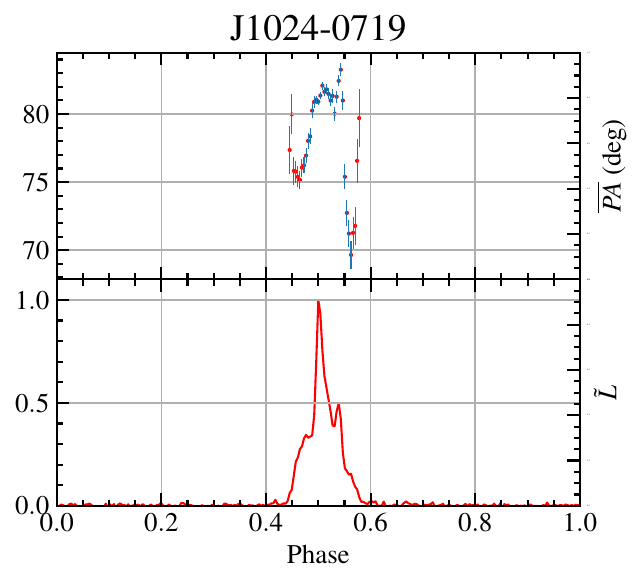}
\includegraphics[width=0.4\textwidth]{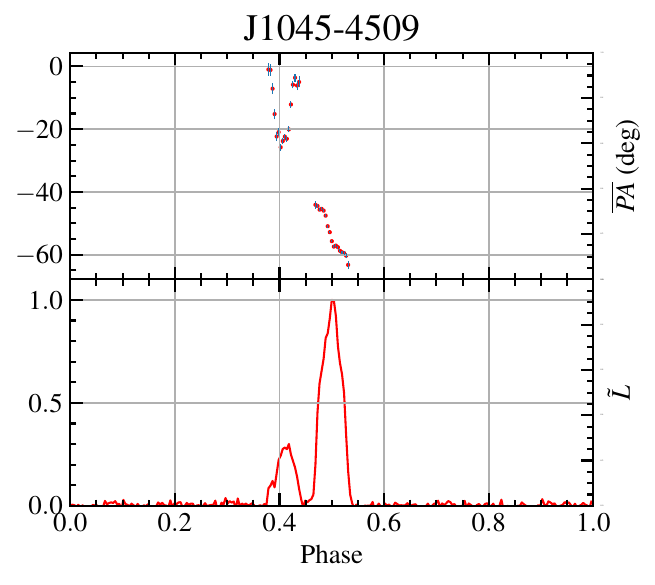}
\includegraphics[width=0.4\textwidth]{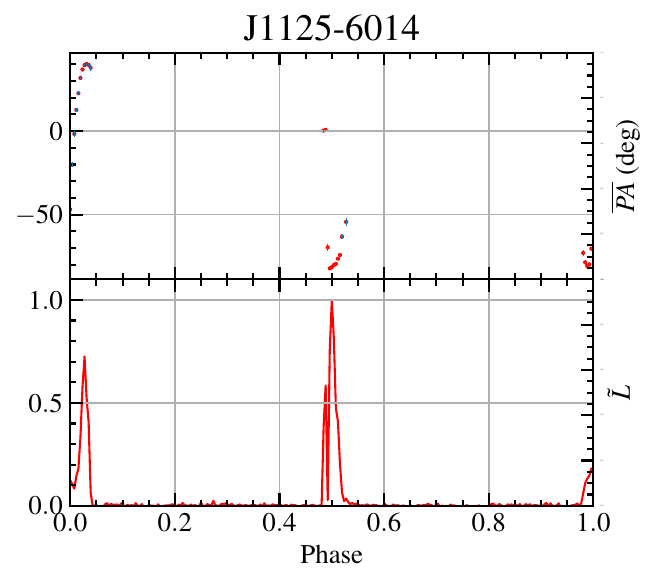}
\includegraphics[width=0.4\textwidth]{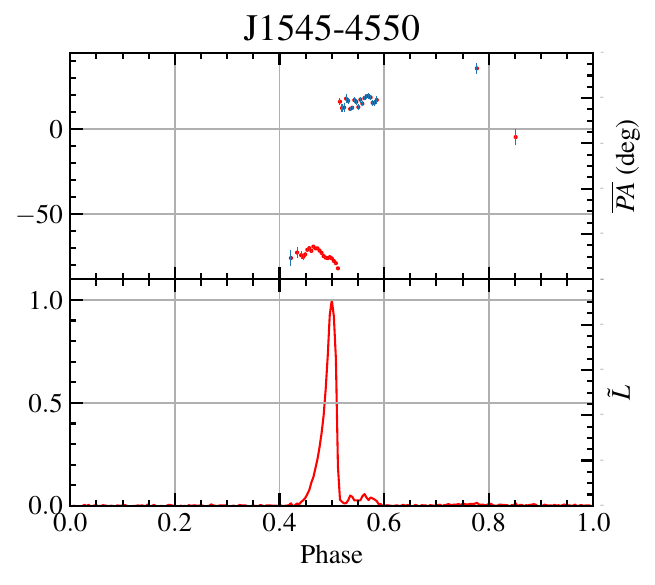}
\includegraphics[width=0.4\textwidth]{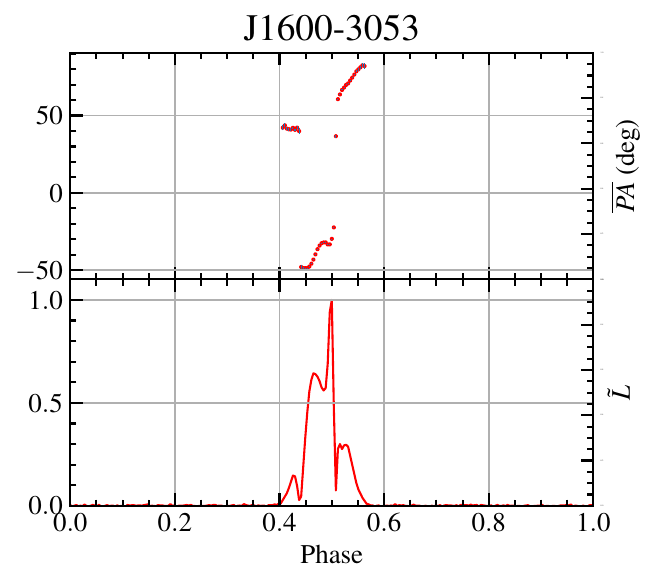}
\includegraphics[width=0.4\textwidth]{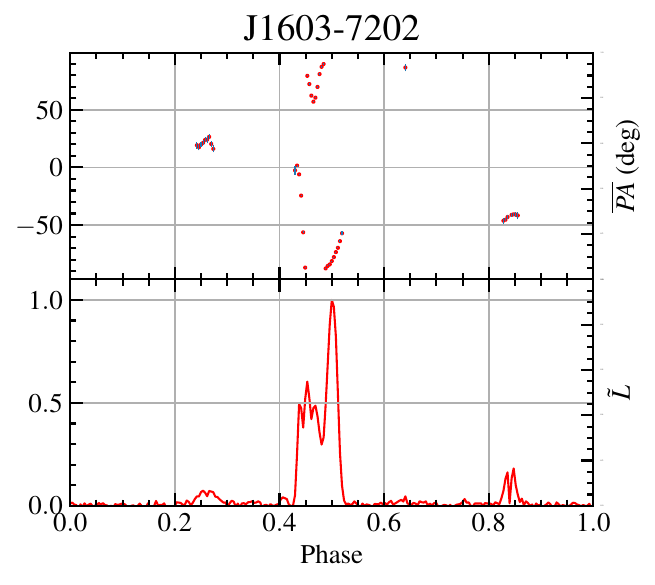}
\caption{Same as Fig.~\ref{fig:Profiles1} for other MSPs.}
\label{fig:Profiles2}
\end{figure*}

\begin{figure*}[ht]
\centering
\includegraphics[width=0.4\textwidth]{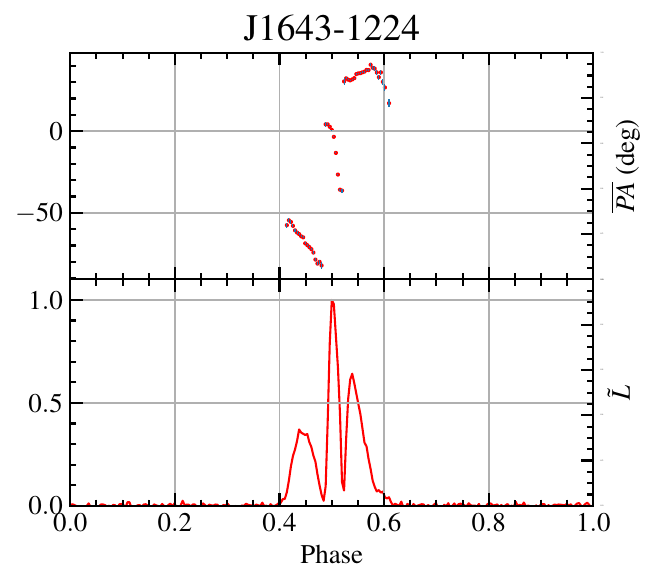}
\includegraphics[width=0.4\textwidth]{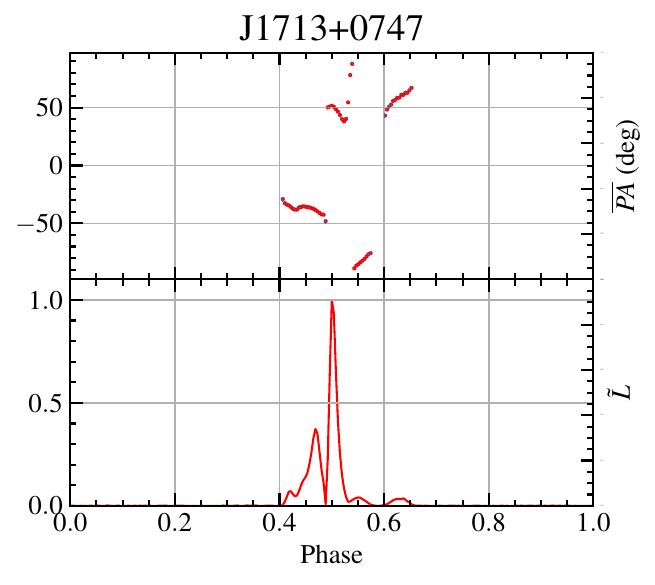}
\includegraphics[width=0.4\textwidth]{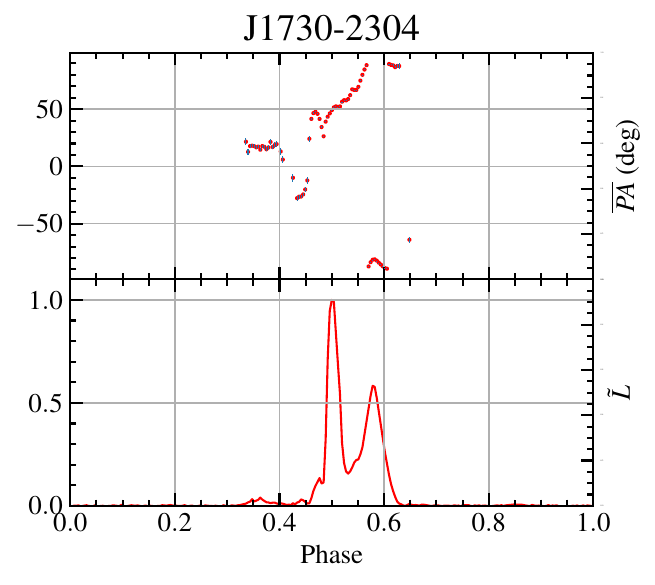}
\includegraphics[width=0.4\textwidth]{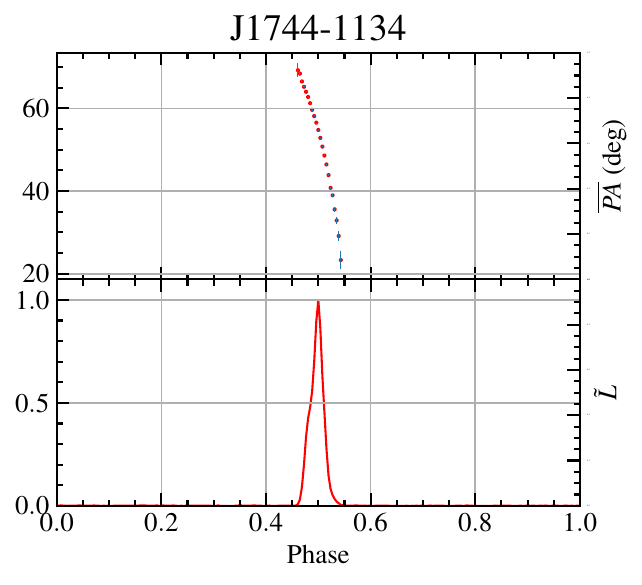}
\includegraphics[width=0.4\textwidth]{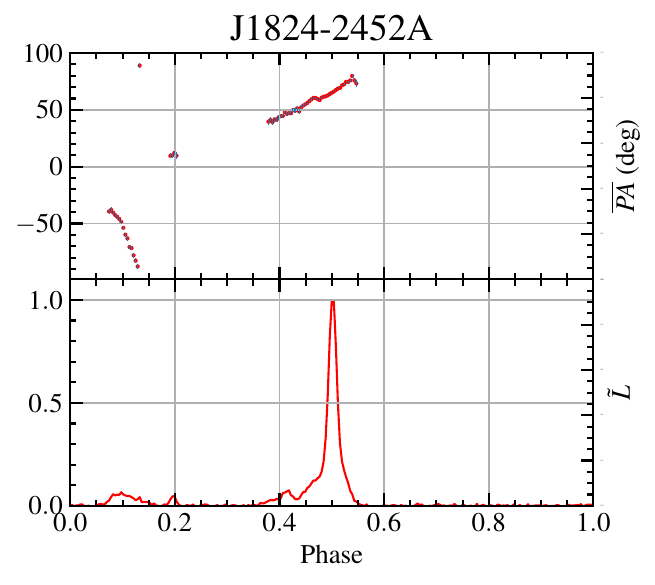}
\includegraphics[width=0.4\textwidth]{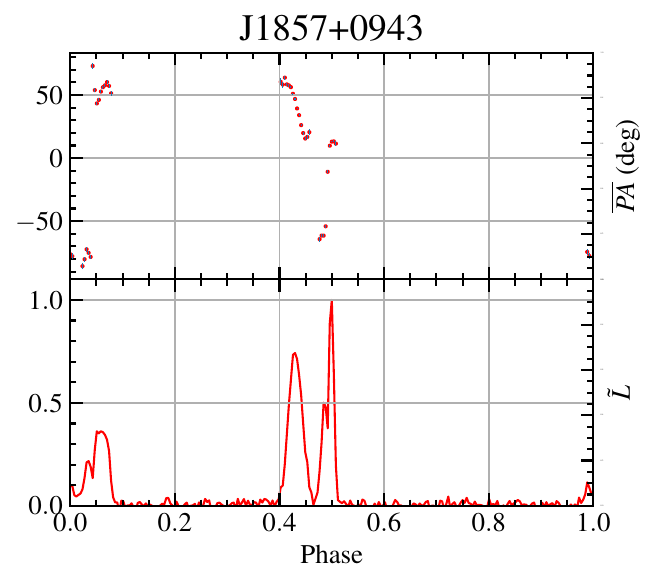}
\caption{Same as Fig.~\ref{fig:Profiles1} for other MSPs.}
\label{fig:Profiles3}
\end{figure*}

\begin{figure*}[ht]
\centering
\includegraphics[width=0.4\textwidth]{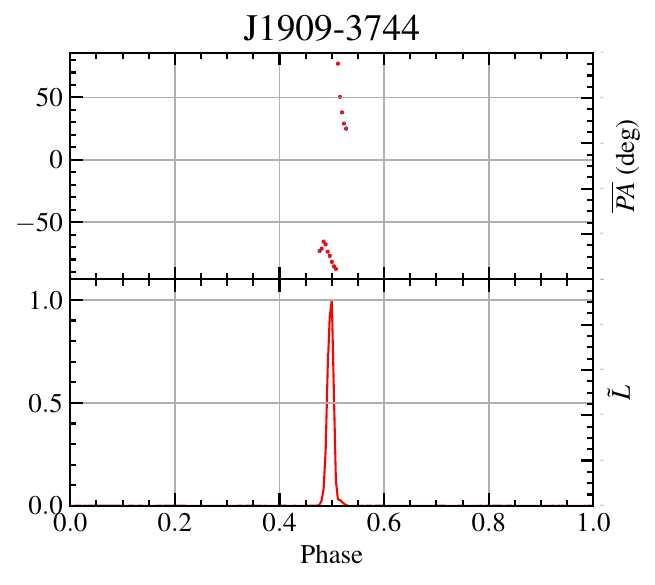}
\includegraphics[width=0.4\textwidth]{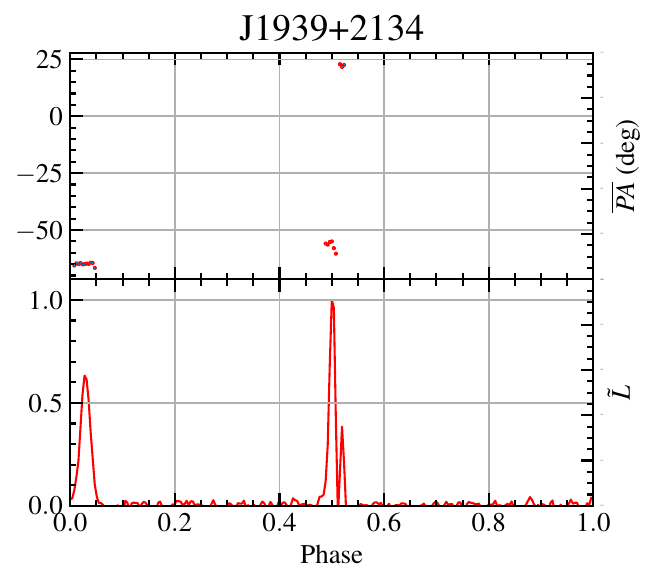}
\includegraphics[width=0.4\textwidth]{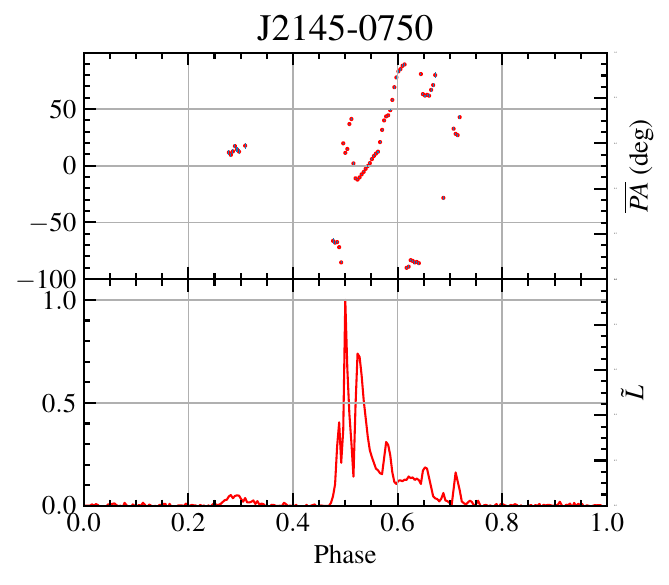}
\includegraphics[width=0.4\textwidth]{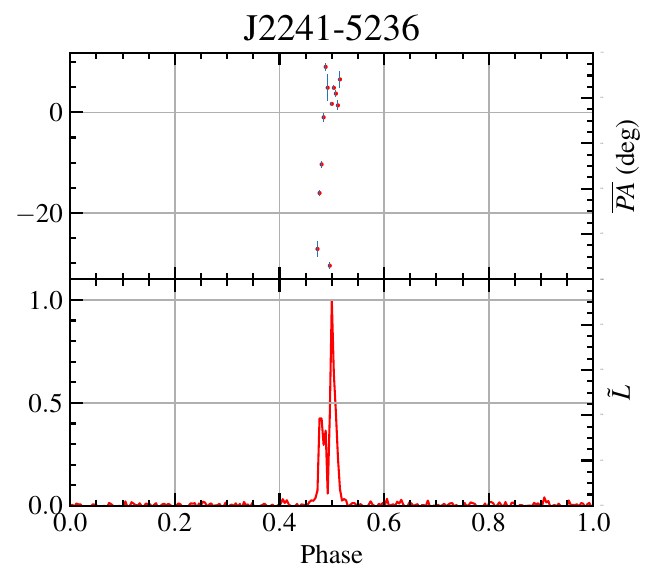}
\caption{Same as Fig.~\ref{fig:Profiles1} for other MSPs.}
\label{fig:Profiles4}
\end{figure*}

Next, we use the data across all epochs to derive the standard profile for the intensity of linear polarization in the phase bins. Following Refs.~\cite{Everett:2000yj,Dai:2015awa}, we first subtract the bias for the measured $L_{n,k}$ at each epoch and phase bin as
\begin{align}
    L'_{n,k} = \begin{cases}
        \sqrt{L_{n,k}^2 - \sigma_{I,n}^2}&L_{n,k}\geq1.57\sigma_{I,n}\\
        0 &L_{n,k}< 1.57 \sigma_{I,n}
    \end{cases},
\end{align}
where $\sigma_{I,n}$ is the off-pulse RMS of the total intensity $I_{n,k}$. We derive the standard profile for the linear polarization intensity as  
\begin{align}
    &\tilde{L}_k\equiv 
\frac{\sum_n w_n L'_{n,k}}{\sum_n w_n},
\end{align}
where $w_n = \max(L'_{n,k}) /\sigma_{S,n}^2$ is weight, with $\max(L'_{n,k})$ being the maximal linear polarization intensity across all phase bins for the $n$-th epoch. We subtracted the baseline value of $\tilde{L}_k$ to focus the subsequent study on the on-pulse bins.

Determining the PA standard profile is less straightforward because it arises from a phase factor. In this analysis, we estimate $\overline{\PA}_k$ for the $k$-th phase bin as 
\begin{equation}
\begin{aligned}
    \overline{\PA}_k
    & \equiv \frac{1}{2}\arctan\left( \frac{\sum_n w_n U_{n,k}}{\sum_n w_n Q_{n,k}} \right) \, .
\end{aligned}
\end{equation}
Figures~\ref{fig:Profiles1}-\ref{fig:Profiles4} display the standard profiles of  $\overline{\PA}_k$ and $\tilde{L}_k$ for the 22 PPPA pulsars. We derive the $\overline{\PA}_k$ errors shown in these figures with an error propagation function. In Table~\ref{tab:pulsar_properties}, we also present the $\max(\tilde{L}_k)/\sigma_{\tilde{L}}$ value for each pulsar, where $\sigma_{\tilde{L}}$ denotes the off-pulse RMS of the standard profile $\tilde{L}_k$, to reflect the sharpness of its pulses.

\subsection{Time series of the PA residuals}

For the mass range of interest, the ALDM signal remains nearly unchanged during a single rotation period of MSPs. However, the standard profiles of intensity and PA of the linearly polarized pulsar light vary significantly across the phase bins due to the complexity of pulsar radiation, as shown in Figures \ref{fig:Profiles1}-\ref{fig:Profiles4}. To reduce such systematic uncertainties in data, we define one PA residual only for each epoch by taking an average over the phase bins.

Concretely, we split first the observed $Q$ and $U$ parameters as:
\begin{equation}
    Q_{n,k} = Q^{\rm t}_{n,k} + n_{n,k} \, , \qquad
    U_{n,k} = U^{\rm t}_{n,k} + n_{n,k}\, ,
\end{equation}
where $n_{n,k}$ is observational noise, with $\langle n_{n,k} \rangle =0$ (here and below $\langle \cdot\cdot\cdot \rangle$ denotes an average over an ensemble of possible realizations for a given pulsar), and 
\begin{eqnarray}
Q^{\rm t}_{n,k} &=& c_n^{\rm t}\tilde{L}_k^{\rm t} \cos\left(2\overline{\PA}_k^{\rm t}+2\Delta \PA_n^{\rm t}\right) \, ,\nonumber\\
U^{\rm t}_{n,k} &=& c_n^{\rm t}\tilde{L}_k^{\rm t} \sin\left(2\overline{\PA}_k^{\rm t}+2\Delta \PA_n^{\rm t}\right) 
\end{eqnarray}
are the true values of $Q$ and $U$, respectively, with $c_n^t$ being a normalization factor. $\Delta \PA_n^{\rm t}$ is the true PA residual for the $n$-th epoch and has no phase bin dependence. 
The observed PA residual is then defined as 
\begin{equation}\label{eq:PAresObs}
\begin{aligned}
    \Delta\PA^{\rm obs}_n 
    &=\frac{1}{2}\arg \mathcal{S}_n   \, ,
\end{aligned}
\end{equation}
with 
\begin{eqnarray}
    \mathcal{S}_n  &=&\sum_k \tilde{L}_k e^{-2i\overline{\PA}_k} \left[Q_{n,k}+iU_{n,k}\right] \nonumber\\
    &=& \sum_k \tilde{L}_kL_{n,k}e^{2i\PA_{n,k}-2i\overline{\PA}_k} \, .
\end{eqnarray}
The measurement error $\sigma_n$ for $\Delta\PA^{\rm obs}_n$ is estimated using the off-pulse standard deviation of the pulse profile and the error propagation formula. This error reflects the radiometer noise for a given observation bandwidth and integration time. 
Here we only consider the phase bins with linear polarization intensities three times greater than the off-pulse RMS. Indeed, we have 
$ \langle \mathcal{S}_n  \rangle =  \mathcal{S}_0c_ne^{2i\PA_n^{\rm t}}$, with $\mathcal{S}_0\equiv\sum_k \tilde{L}_k^{\rm t}\tilde{L}_k e^{ 2i\overline{\PA}_k^{\rm t} - 2i\overline{\PA}_k }$, and hence  
\begin{eqnarray}
\langle  \Delta\PA^{\rm obs}_n  \rangle = \Delta \PA_n^{\rm t} \, .
\end{eqnarray}
The constructed time series of PA residual thus provide an unbiased signal estimation.

\begin{figure*}[ht]
\centering
\includegraphics[width=0.92\textwidth]{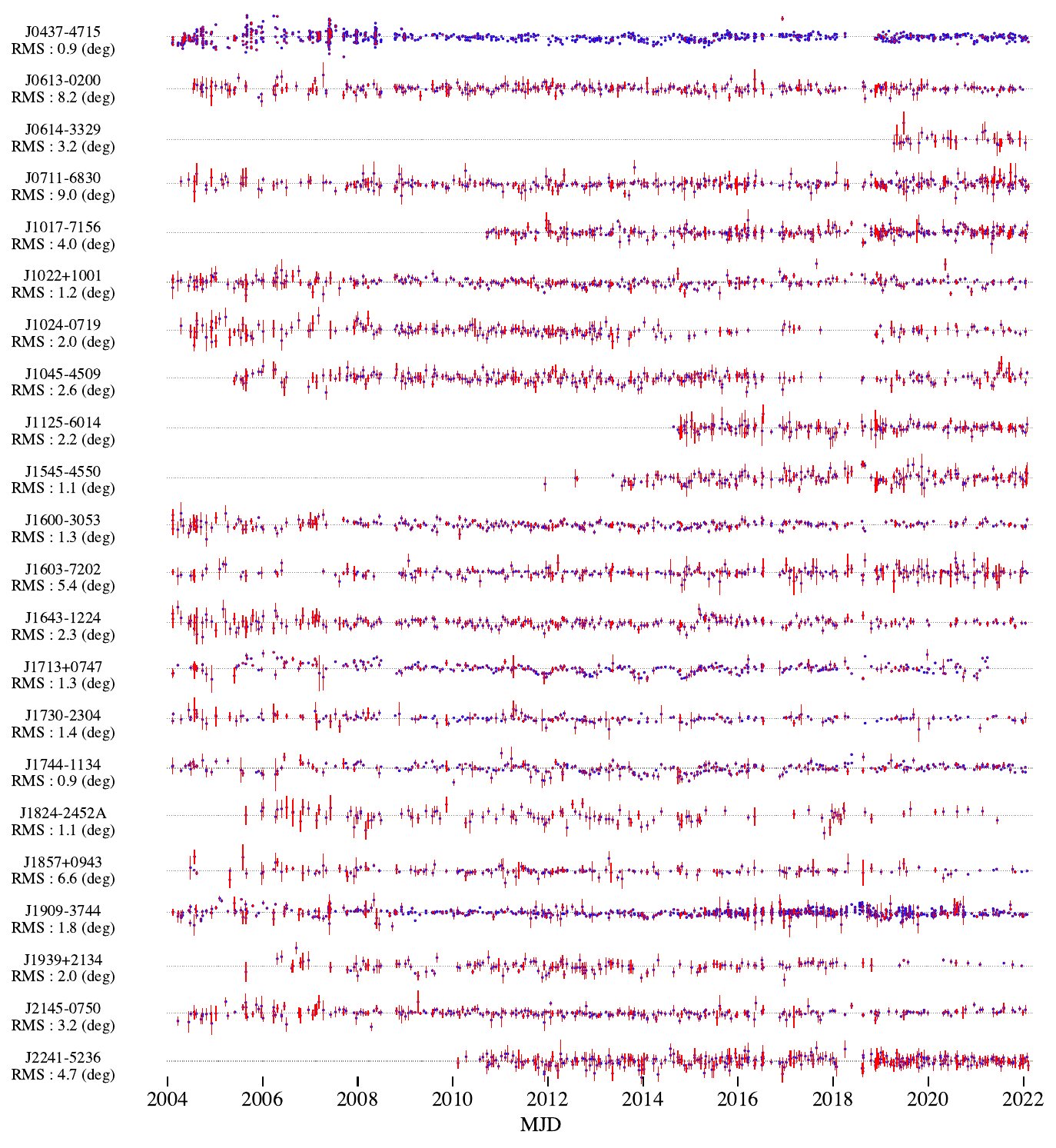}
\caption{Time series of the observed PA residuals in Eq.~(\ref{eq:PAresObs}) for the 22 PPPA MSPs. 
The central value of each observation and its error bar are denoted in blue and red, respectively. }
\label{fig:dPA}
\end{figure*}

\begin{figure*}[ht]
\centering
\includegraphics[width=0.92\textwidth]{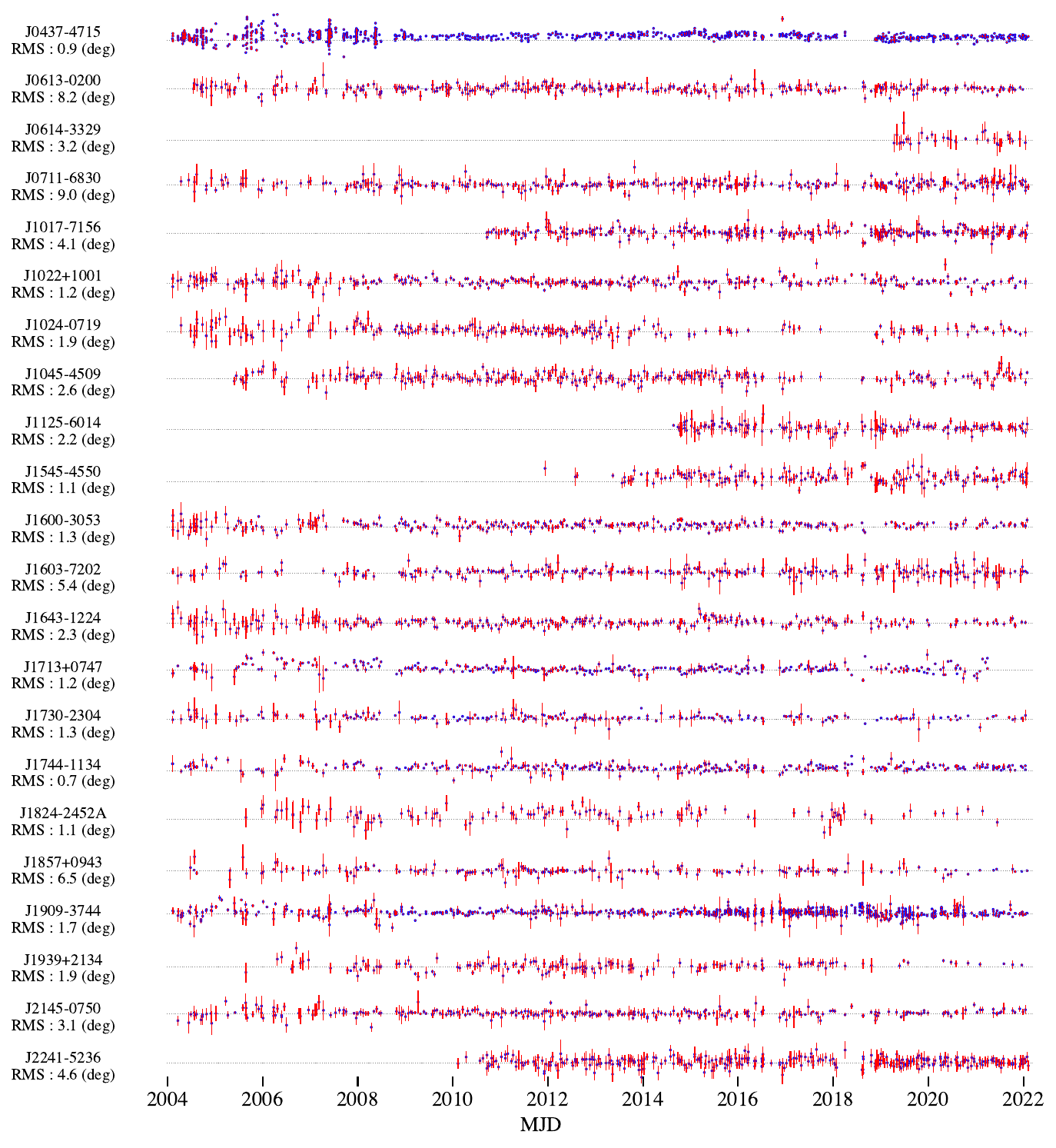}
\caption{Time series of the observed PA residuals in Eq.~(\ref{eq:PAresObs}) for the 22 PPPA MSPs,  after the ionospheric subtraction with \texttt{ionFR}. The central value of each observation and its error bar are denoted in blue and red, respectively.}
\label{fig:dPA_ionfr}
\end{figure*}

Fig.~\ref{fig:dPA} displays the time series of PA residuals for the 22 PPPA MSPs. In the analysis, we discard a small number of observations where the PA residuals deviate by more than $\pm 4\sigma$ from the baseline or have an error bar larger than $4\sigma$ or $30$ degrees. These observations account for less than 5\% of total observations for all pulsars. One observation in Fig.~\ref{fig:dPA} is the annual oscillations in $\Delta \PA^{\rm obs}$ for pulsars such as PSRs~J1713+0747 and J1744$-$1134. These oscillations are likely driven by variations in ionospheric total-electron content~\cite{Sotomayor:2013}, rather than by solar elongation. Actually, one can see from Table~\ref{tab:pulsar_properties} that PSRs~J1713+0747 and J1744$-$1134 have ecliptic latitudes of 30.7$^\circ$ and 11.8$^\circ$, respectively, much higher than those of many other pulsars in our sample which do not show such a time-varying trend.  In contrast, J1022+1001 has the lowest ecliptic latitude in our sample at $-0.1^\circ$, yet showing no significant annual variation. A detailed discussion on the ionospheric effects and the noises that they may  introduce on various timescales is provided in Sec.~B.

\subsection*{Pulsar distance}

Pulsar distances are treated as free parameters in the analysis, with priors determined by their respective measurement methods. For most pulsars, their distances to the Earth have been measured with parallax (PX), where 
\begin{align}
    \frac{D}{\rm kpc} = \left(\frac{\PX}{\rm mas}\right)^{-1} \, .
\end{align}
The PX distance has a prior   
    	\begin{align}
		&\ln P\left(D^*\right) =  -
  \frac{1}{2}\frac{(1/D^*-1)^2}{\sigma_{D^*}^2} - 2\ln D^*+ c \, ,\label{eq:pxg}  
	\end{align}
with  
\begin{align}
  \sigma_{D^*} \equiv \frac{D_0}{\rm kpc}\frac{\sigma_{\PX}}{\rm mas} \, ,\qquad D^* \equiv D/D_0 \, .
\end{align}
Here $D_0$ is the central value of the PX distance. If the error $\sigma_{\PX}$ is less than 30\% of the PX itself~\cite{arzoumanian2023nanograv}, we take the PX distance for the analysis. Otherwise, we estimate the pulsar distance with another method - dispersion measure - using the Galactic electron density model~\cite{yao2017new}. The dispersion-measure ($\widetilde {\rm DM}$) distance has a prior of half-Gaussian distribution:
	\begin{align}
		\ln P\left(D^*\right)=c -\begin{cases}
			\frac{1}{2\sigma_{D^*}^2}(D^*-0.8)^2,&0< D^*\leq0.8\\
			0,&0.8< D^*\leq 1.2\\
		\frac{1}{2\sigma_{D^*}^2}(D^*-1.2)^2,&D^*>1.2\\
		\end{cases},\label{eq:uhg}
	\end{align}
where $\sigma_{D^*}=0.2$. Accordingly, $D_0$ becomes the central value of the $\widetilde{\rm DM}$ distance. In both cases, we determine the constant in the prior by a numerical integration to fulfill the normalization condition. The $\widetilde{\rm DM}$ and PX distances, along with their respective errors, for the PPPA pulsars listed in Table~\ref{tab:pulsar_properties}. Note that we also need the right ascension and declination of pulsars in the signal covariance matrix for calculating the mutual distances between the arrayed pulsars. However, since these coordinates have been measured with negligible uncertainties, we do not list them in Table~\ref{tab:pulsar_properties}. Pulsar ephemerides used in this work are all taken from the ATNF pulsar catalogue~\cite{Manchester:2004bp}.

\section{B. Bayesian Analysis Framework}
\label{sec:Bayes}

This framework includes the modeling of PA residuals and the Bayesian analysis of detection sensitivities. We model the observed PA residuals as described in Eq.~(\ref{eq:PAmodel}), by summing up the contributions from various noises and the ALDM signal. Below, we will discuss these two components one by one.

\subsection*{Noise modeling}

One major contribution to the observed PA residuals is Faraday rotation induced by the Earth's ionosphere. This effect can be described as $\Delta {\rm PA}^{\rm ion} = \phi^{\rm ion}\lambda^2$, and is important for radio band. Here $\phi^{\rm ion}$ is ionospheric rotation measure (RM). For a long time, the ionospheric effects have been calculated with the numerical code \texttt{FARROT}, developed at the Dominion Radio Astrophysical Observatory, and the International Reference Ionosphere (\texttt{IRI}) model~\cite{Han:2006ci}. 
We instead employ \texttt{ionFR}~\cite{Sotomayor:2013}, a program developed more recently to calculate the ionospheric RM using the International Geomagnetic Reference Field (IGRF-13) model and publicly available global electron maps. We provide the coordinates and observation epochs of the PPPA pulsars and the Earth's position of Parkes observatory as inputs to \texttt{ionFR} to determine the line-of-sight to each observation and compute its ionospheric RM. As the ionospheric data is available on an hourly basis, a binning approach is applied for this calculation. Concretely, if the median time of an observation epoch is between 30 minutes before and after an exact hour, we subtract the observed PA residual by the ionospheric RM of this hour. This hourly timescale sets a cutoff for the sensitivity range at the large mass end. In Fig.~\ref{fig:dPA_ionfr}, we demonstrate  the time series of PA residuals for the 22 PPPA MSPs, after the ionospheric subtraction with \texttt{ionFR}.

The ionospheric Faraday rotation is strongly influenced by solar activities, and its contributions to the observed PA residuals may vary significantly over time, on scales of hours to years, and direction. 
For example, solar activities can greatly impact the ionospheric total electron content, leading to considerable annual variations in the observed PA residuals. As illustrated in Fig.~\ref{fig:dPA_ionfr}, the apparent annual oscillations for PSRs J1713+0747 and J1744$-$1134 in Fig.~\ref{fig:dPA} get significantly reduced - though not entirely eliminated - supporting the conjecture of their ionospheric origin~\footnote { The observed annual oscillations for these pulsars may receive contributions also from the ISM effects~\cite{hotan2004psrchive,Everett:2000yj,arzoumanian2023nanograv}, such as the annual variations of magnetic field strength caused by magnetized clumps of material along the line of sight to the pulsars~\cite{arzoumanian2023nanograv}.}. Generally, the complexity of solar activities make it challenging to predict the ionospheric effects and hence yield a large uncertainty for the predicted RM values. 
The \texttt{ionFR} prediction for a given pulsar may fluctuate by up to $\sim1$\,rad\,m$^{-2}$ within two hours, particularly during intense solar activities, and can also differ from the \texttt{IRI} one by $\mathcal O(1)$ rad\,m$^{-2}$. 
Furthermore, for many pulsars, we observe a discrepancy 
between the median measurement error and the SD of PA residuals after ionospheric subtraction (see the last two columns in Table~\ref{tab:pulsar_properties}). This suggests the presence of unmodeled noise in the PA residual time series, despite the efforts to model ionospheric corrections using \texttt{ionFR}. To account for these observations, we incorporate additional noise
components alongside radiometer noise and modeled ionospheric corrections, following the common practices in the PTA analysis~\cite{Lentati:2014}.
This advanced noise modeling has not been considered in the existing analyses of pulsar polarization data, and turns out to be crucial for optimizing the PPA detection of the ALDM signal.

The random noise includes temporally uncorrelated (white noise) and correlated (red noise) parts. Our modeling of white noise is similar to that of the PTA analysis, with the covariance matrix
\begin{align}
    C^w_{p,n;q,m}=(\sigma_{p,n}^{\prime})^2\delta_{p,q}\delta_{n,m},\; 
    (\sigma_{p,n}^{\prime})^2 =({\rm EF}_p\,\sigma_{p,n})^2  + {\rm EQ}_p^2\,.
    \label{eq:new_sigma}
\end{align}
Here $\sigma_{p,n}$ is the measurement error of the observed PA residual $\Delta\PA^{\rm obs}_{p,n}$ for the $n$-th epoch of the $p$-th pulsar, EF$_p$ is a rescaling factor of the measurement error, and EQ$_p$ is additional white noise added in quadrature. EF$_p$ and EQ$_p$ are free parameters. We introduce them to collectively account for the contributions from uncorrelated and weakly correlated processes, considering the complexity of measuring the individual components of white noise.
Their prior distributions are given by
\begin{eqnarray}
P\left(\log_{10}({{\rm EF}_p
    })\right) &=& {\rm U}[-2,2]\, ,\nonumber\\ 
    P\left(\log_{10}({{\rm EQ}_p
    }/{\rm rad})\right) &=& {\rm U}[-8,2]\, ,
\end{eqnarray}
where ${\rm U}$ is a uniform distribution. 
Note, EQ$_p$ has been commonly used for modeling jitter noise, namely random pulse-to-pulse variation, in high-precision pulsar timing analysis~\cite{sod+14}. For the pulsar polarization data analyzed here, the jitter noise is not significant as a contribution to the observed PA residuals. Actually, the measured jitter noise in the considered pulsars yields a profile variation by only up to $\sim 1/100$ of a phase bin. However, complex effects such as underestimated or overestimated error bars, $\Delta \PA$ variations at short timescales due to unmodeled ionospheric effects, environmental fluctuations, and even residual RFI effects after the mitigation procedure~\cite{2023PASA...40...49Z}  could bring in additional white noise and contribute to EF$_p$ and EQ$_p$.

For the red noise, we only consider the contributions uncorrelated between different pulsars, such as the RM variations caused by the temporal effects in the ISM~\cite{smt+23}. The covariance matrix is then block-diagonal, represented as $\bC^r = {\rm diag}\{C_1^r,\,\dots,C_{\mathcal{N}}^r\}$, where $\mathcal{N}$ denotes the number of pulsars.
We model the red noise with a power-law spectrum in the frequency domain. The covariance matrix in the time domain is thus given by
\begin{align}
    \left(C_{p}^r\right)_{n,m} = (S_{p}^r)^2
    \int_{f_{L}}^{f_{H}} \frac{\dd f}{f_{\rm yr}} \left( \frac{ f}{f_{\rm yr}} \right)^{\Gamma_p}\cos\left[2\pi f( t_{p,n} - t_{p,m}) \right] \, ,\label{eq:red_noise_continuous}
\end{align}
where $f_L$ and $f_H$ are the low- and high-frequency cutoffs, and $S_{p}^r$ and $\Gamma_p$ have priors 
\begin{equation}
\begin{aligned}
    P\left(\log_{10}({S_{p}^r}/{\rm rad})\right) = {\rm U}[-8,2]  \, ,\quad 
    P\left(\Gamma_p\right) = {\rm U}[-8,2]   \, .
\end{aligned}
\end{equation} 
We choose $f_L=1/T_{{\rm obs},p}$, with $T_{{\rm obs},p}$ being the observation period for the $p$-th pulsar, and further discretize this integration as   
\begin{eqnarray}
        \left(C_{p}^r\right)_{n,m} &=&  (S_{p}^r )^2
        \sum_{k=1}^{k_{{\rm max},p}}  \frac{1 }{f_{\rm yr}T_p} \left( \frac{ k}{f_{\rm yr}T_p} \right)^{\Gamma_p}\nonumber\\
        &&\cos\left[\frac{2\pi k}{T_p}( t_{p,n} - t_{p,m}) \right]\, ,\label{eq:red_noise_discrete}
\end{eqnarray}
with a spacing of $\Delta f = 1/T_{{\rm obs},p}$. The approach to address red noise here is similar to that used for the PTA studies~\cite{agazie2023nanograv}. The major source could be the ISM where the RM varies linearly over time with a slope $\sim 0.1 - 1$\,rad\,m$^{-2}$ per year, correlated with $\widetilde {\rm DM}$~\cite{kjk+24}. Additionally, unmodeled ionospheric effects after the  \texttt{ionFR} subtraction may also vary over time and contribute to the red noise for some pulsars.

 
To account for potential DM and RM variations of long timescale, we incorporate a polynomial component of up to second-order in the noise model:
\begin{align}
    \Delta\PA^{\rm pol}_p(t) = \psi^{(0)}_p + \psi^{(1)}_p t + \psi^{(2)}_p t^2 \, ,
\end{align}
where $\psi^{(0)}_p, \psi^{(1)}_p, \psi^{(2)}_p$ are the coefficients of intrinsic, linear and quadratic terms, respectively. These temporal variations could arise from slow changes in ISM properties, particularly due to gradual stochastic variations in electron density included in the DM and RM~\cite{2016ApJ...821...66L,kjk+24}. For a given pulsar, $\Delta\PA^{\rm pol}_p(t)$ can be recast as a vector of its values at the relevant observation epochs $\{ t_{p,n} \}$: 
\begin{align}
    \Delta\bPA^{\rm pol}_p = \bM_p \bPsi_p
\end{align}
with
\begin{align}
 \bM_p = \begin{pmatrix}
			1 & t_{p,1} & t_{p,1}^2 \\
			1 & t_{p,2} & t_{p,2}^2 \\
			\vdots & \vdots & \vdots \\
            1 & t_{p,N_p} & t_{p,N_p}^2
		\end{pmatrix}, \qquad \bPsi_p = \begin{pmatrix} \psi^{(0)}_p & \psi^{(1)}_p & \psi^{(2)}_p \end{pmatrix}^T   \, .
\end{align}
This can be generalized to the case of  $\mathcal{N}$ pulsars, yielding 
\begin{eqnarray}
    \Delta\bPA^{\rm pol} &=& \begin{pmatrix}
        \Delta\bPA^{\rm pol}_1 \\ \Delta\bPA^{\rm pol}_2 \\ \vdots \\ \Delta\bPA^{\rm pol}_\mathcal{N} \end{pmatrix}, \quad \bPsi = \begin{pmatrix}
        \bPsi_1 \\ \bPsi_2 \\ \vdots \\ \bPsi_\mathcal{N} \end{pmatrix}, \nonumber\\ 
        \bM &=& \begin{pmatrix}
            \bM_1 & \mathbf{0} & \cdots & \mathbf{0} \\
            \mathbf{0} & \bM_2 & \cdots & \mathbf{0} \\
            \vdots & \vdots & \ddots & \vdots \\
            \mathbf{0} & \mathbf{0} & \cdots & \bM_\mathcal{N}
        \end{pmatrix}  \, ,
\end{eqnarray}
Now plugging these terms in Eq.~\eqref{Eq:likelihood}, we have
\begin{eqnarray}
     \ln\mathcal{L} &=& - \frac{1}{2} \bPsi^T \bM^T \bC^{-1} \bM \bPsi + \delta\bPA^T \bC^{-1}\bM\bPsi \nonumber\\
     &&- \frac{1}{2} \delta\bPA^T \bC^{-1}\delta\bPA - \frac{1}{2}\ln |2\pi\bC|\, ,
\end{eqnarray}
where $\delta \PA=\Delta\PA^{\rm obs}-\Delta\PA^{\rm ion}$. By integrating out $\bPsi$ analytically, we finally obtain a marginalized likelihood
\begin{eqnarray}\label{Eq:Mlikelihood}
    \ln\mathcal{L}_m &\propto& \frac{1}{2} \left( \delta \bPA^c \right)^T \bC_M^{-1} \delta \bPA^c - \dfrac{1}{2}\ln |2\pi\bC_M| \nonumber\\
    &&- \frac{1}{2} \delta\bPA^T \bC^{-1}\delta\bPA - \frac{1}{2}\ln |2\pi\bC|,
\end{eqnarray}
with $\bC_M = \bM^T \bC^{-1} \bM$ and $(\delta\bPA^c)^T = \delta\bPA^T \bC^{-1}\bM$.

\begin{table*}[t]
    \centering
	\begin{tabular}{|c|c|cc|cc|cccc|c|}
\hline
        & \multirow{2}{*}{Pulsar}  & \multicolumn{4}{c|}{Posterior} &
        \multirow{2}*{$\ln{\rm BF}^{\psi^{(0,1)}}_{\psi^{(0)}}$}&
        \multirow{2}*{$\,^{\psi^{(0,1,2)}}_{\psi^{(0,1)}}$}&
        \multirow{2}*{$\,^{\psi^{(0,1,2)}+w}_{\psi^{(0,1,2)}}$}& \multirow{2}*{$\,^{\psi^{(0,1,2)} +w+r}_{\psi^{(0,1,2)}+w}$}  \\
        \cline{3-6}
		 &  & $\log_{10}$EF$_p$ & $\log_{10}$EQ$_p$/deg & $\log_{10}S^{\rm r}_p$/deg &$\Gamma_p$ &&&& \\
   \hline
        \hline
1 & J0437--4715 & $0.69^{+0.03}_{-0.03}$  & $-0.65^{+0.16}_{-3.31} $ & $ -0.92^{+0.12}_{-0.18}$ & $-1.27^{+0.39}_{-0.49}$ & 399.4 & 326.7 & 12649.8 & 30.2\\
2 & J0613--0200 & $-0.19^{+0.05}_{-0.06}$  & $ 0.72^{+0.05}_{-0.06} $ & \dots & \dots & -2.5 & 1.1 & 12.5 & -0.1\\
3 & J0614--3329 & $-0.18^{+0.06}_{-0.06}$  & \dots & \dots & \dots & -2.4 & -0.8 & 2.4 & -0.1\\
4 & J0711--6830 & $-0.09^{+0.03}_{-0.03}$  & $ 0.51^{+0.05}_{-0.06} $ & \dots & \dots & -3.0 & -1.6 & 18.4 & -0.4\\
5 & J1017-7156 & $0.03^{+0.03}_{-0.04}$  & $-0.14^{+0.30}_{-3.90} $ & \dots & \dots & -3.2 & -2.6 & 5.2 & -0.1\\
\hline
6 & J1022+1001 & $-0.05^{+0.04}_{-0.04}$  & $-0.22^{+0.06}_{-0.07} $ & $ -1.62^{+0.83}_{-2.96}$ & $-1.67^{+1.22}_{-2.84}$ & -3.1 & -3.2 & 22.8 & 0.8\\
7 & J1024--0719 & $-0.07^{+0.02}_{-0.03}$  & \dots & \dots & \dots & -3.2 & -1.8 & 2.9 & -0.1\\
8 & J1045--4509 & $-0.03^{+0.02}_{-0.02}$  & \dots & \dots & \dots & -3.5 & -2.1 & 0.1 & -0.1\\
9 & J1125--6014 & $-0.17^{+0.03}_{-0.03}$  & \dots & \dots & \dots & -3.6 & -0.7 & 14.7 & -0.2\\
10 & J1545--4550 & $-0.20^{+0.15}_{-0.16}$  & $-0.21^{+0.10}_{-0.38} $ & $ -0.48^{+0.08}_{-0.08}$ & $-0.05^{+0.61}_{-0.54}$ & -2.6 & -2.8 & 24.6 & 6.5\\
\hline
11 & J1600--3053 & $-0.08^{+0.05}_{-0.05}$  & $-0.10^{+0.05}_{-0.06} $ & \dots & \dots & -4.0 & -3.4 & 17.9 & -0.3\\
12 & J1603--7202 & $-0.07^{+0.02}_{-0.02}$  & $-0.07^{+0.11}_{-0.20} $ & \dots & \dots & -1.5 & -2.3 & 4.5 & -0.1\\
13 & J1643--1224 & $-0.04^{+0.02}_{-0.02}$  & \dots & \dots & $-1.33^{+1.64}_{-3.34}$ & -3.7 & -2.5 & -0.2 & 0.3\\
14 & J1713+0747 & $-0.10^{+0.06}_{-0.07}$  & $-0.06^{+0.03}_{-0.03} $ & $ -2.13^{+0.80}_{-0.84}$ & $-4.59^{+1.67}_{-1.71}$ & 200.6 & 115.6 & 924.7 & 6.6\\
15 & J1730--2304 & $-0.17^{+0.04}_{-0.04}$  & $-0.15^{+0.04}_{-0.04} $ & \dots & \dots & -5.1 & -0.9 & 126.6 & -0.1\\
\hline
16 & J1744--1134 & $0.01^{+0.03}_{-0.04}$  & $-0.42^{+0.04}_{-0.05} $ & $ -1.87^{+1.00}_{-3.07}$ & $-0.97^{+1.21}_{-2.61}$ & -5.2 & -4.1 & 227.8 & 0.4\\
17 & J1824--2452 & $-0.06^{+0.03}_{-0.03}$  & \dots & \dots & \dots & -4.1 & -2.3 & 0.0 & -0.0\\
18 & J1857+0943 & $-0.10^{+0.03}_{-0.03}$  & \dots & \dots & \dots & -1.9 & -1.0 & 5.1 & -0.1\\
19 & J1909--3744 & $-0.04^{+0.03}_{-0.03}$  & $ 0.00^{+0.02}_{-0.02} $ & $ -0.25^{+0.06}_{-0.06}$ & $-0.39^{+0.34}_{-0.34}$ & 49.4 & 27.9 & 3735.3 & 53.8\\
20 & J1939+2134 & $-0.20^{+0.06}_{-0.09}$  & $ 0.08^{+0.08}_{-0.08} $ & \dots & \dots & 15.3 & 1.5 & 16.8 & -0.3\\
\hline
21 & J2145--0750 & $-0.08^{+0.02}_{-0.02}$  & \dots & \dots & \dots & -3.2 & 0.7 & 5.2 & 0.1\\
22 & J2241--5236 & $-0.10^{+0.02}_{-0.02}$  & \dots & \dots & \dots & -3.3 & -1.8 & 12.9 & -0.0\\
\hline
		\hline
        \end{tabular}
    \caption{Performance analysis of different noise models for the 22 PPPA MSPs, using the data after ionospheric subtraction. The four ``Posterior'' columns display the posterior distributions of the white-noise and red-noise parameters in the full noise model, where the unconstrained parameters are marked as ``\dots". The last four columns show the Bayes factors for model comparison. Here, $\psi^{(0)}$,  $\psi^{(0,1)}$, and $\psi^{(0,1,2)}$ denote the noise models with the minimal white noise of $\sigma_{p,n}$ for $\Delta\PA^{\rm obs}$ and the polynomial noise term up to the $0^{\rm th}$, $1^{\rm st}$, and $2^{\rm nd}$ orders, respectively. The notations of ``$+w$'' and ``$+r$'' indicate the incorporation of additional white-noise and red-noise components in the noise model, respectively. }
    \label{tab:pulsar_noise}
\end{table*}

We then perform a Bayesian analysis with $\mathcal{L}_m$ to examine the performance of the noise models, by fitting them to the data of 22 PPPA MSPs after ionospheric subtraction using \texttt{ionFR}. The results are summarized in Table~\ref{tab:pulsar_noise}. 
To assess model performance, we compute Bayes factors $\textrm{BF}_{j}^i=Z_i/Z_j$, where $Z_i$ represents the evidence of the $i$-th model and is derived using the Markov Chain Monte Carlo (MCMC) chains generated by the parallel tempering MCMC package \texttt{PTMCMCSampler}~\cite{justin_ellis_2017_1037579} with thermodynamical integral~\cite{goggans2004using, calderhead2009estimating}.
We demonstrate the improvement of noise modeling (indicated by a positive Bayes factor) by including additional components one by one into the minimal noise model $\psi^{(0)}$, which contains the minimal white noise (with EF$_p =1$ and EQ$_p=0$)  and a constant offset ({\it i.e.}, the $0^{\rm th}$-order 
polynomial noise term) 
only, ultimately leading to the full noise model. The results are summarized in the last four columns of Table~\ref{tab:pulsar_noise}. 

Evidently, the additional components (or at least one of them) in the full noise model (see Eq.~(\ref{eq:PAmodel})) improve the data fit for most of the 22 pulsars, likely excluding PSRs J1045--4509, J1643--1224, and J1824--2452, when compared to the minimal noise model. 
The improvement is especially significant for pulsars such as J0437--4715, J1713+0747, and J1909--3744. The relatively narrow posterior distributions for the random noise parameters in the full noise model also reflect the reasonableness of this model. The contribution of each new component varies for different pulsars. For most pulsars, the additional white noise component leads to noticeable improvements, suggesting the possible presence of ionospheric Faraday rotation, caused by imperfect modeling, on timescales shorter than the observation cadence. Several pulsars show evidence or strong hints of red noise, including J0437--4715, J1022--1001, J1545--4550, and J1909--3744. Additionally, despite the reduction of their annual modulation in PA residuals through ionospheric subtraction, J1713+0747 and J1744--1134 also exhibit this characteristic. The ISM variations may contribute significantly to this feature, particularly for PSR J1545$-$4550 which has the highest-$\widetilde {\rm DM}$ value among these six pulsars (see Table~\ref{tab:pulsar_properties}). Since we primarily consider the long-term effects by setting $ f_H \leq 30 f_L $ with $f_L = 1 / T_{{\rm obs},p} $ for the red-noise model given in Eq.~(\ref{eq:red_noise_continuous}), the timescales for the observed red noise might be years-long. This conjecture gets supported by our examination of the $\Delta \PA^{\rm obs}$ structure functions for these pulsars,  using a method from~\cite{Kumamoto:2020oqi}. There is no  power-law rise indicated in these functions. However, we did observe features over timescales of several years. These timescales exceed typical timescales of diffractive scintillation in the ISM~\cite{2013MNRAS.429.2161K}, and fall within the range associated with ISM refractive scintillation~\cite{Kumamoto:2020oqi} and the evolution of pulsar RM~\cite{kjk+24}. Therefore, the observed red noise may arise from ISM structure and turbulence or fluctuation on these scales, as well as variations in the interstellar magnetic field along the line of sight to these pulsars.
Notably, the relatively low $ \widetilde {\rm DM} $ and RM values ($|\rm{RM}|<10$ rad m$^{-2}$) for the pulsars such as J0437--4715 and J1909--3744 may suggest that the observed red noise also receives contributions from the unmodeled ionospheric effects over the years.
Finally, several pulsars exhibit strong evidence of linear and quadratic variations over long periods, as additional noise contributions. Similar trends have been observed also in the DM and RM evolution of pulsars~\cite{kjk+24,2016ApJ...821...66L}.These outcomes reasonably justify our noise modeling.

We will utilize this advanced noise model for the data analysis generally. However, including red noise in the combined sensitivity analysis can significantly increases the numerical cost. The larger $k_{{\rm max},p}$ is, the longer it takes to calculate the inverse of the covariance matrix. So, we choose $k_{{\rm max},p}=30$ to accurately determine the red noise parameters $\Gamma_p$ and $S_p^r$ in the single pulsar analysis, and include the red noise component only for the pulsars with $\ln{\rm BF}^{\psi^{(0,1,2)} +w+r}_{\psi^{(0,1,2)} +w} >2.3$, namely J0437--4715, J1545--4550, J1713+0747 and J1909--3744, in the PPPA analysis. In addition, in the combined analysis, we choose $k_{{\rm max},p}$ according to the best-fit $\Gamma_p$ value from the single pulsar noise analysis. Specifically, we choose $k_{{\rm max},p}=5$ if $\Gamma_p<-3$, and $k_{{\rm max},p}=30$ if $\Gamma_p\geq-3$ to ensure that Eq.~(\ref{eq:red_noise_discrete}) converges to  Eq.~(\ref{eq:red_noise_continuous}) sufficiently.

\subsection*{Stochastic spacetime dependence of the ALDM field}
\label{sec:ALDMfield}

The ALDM field is described in Eq.~(\ref{eq:axionf}) as a random superposition of numerous uncorrelated particle plane waves. This results in a stochastic spacetime dependence in $a(\mathbf{x},t)$ for a given realization of the random parameters  $\{\alpha_{\mathbf{v}},\phi_{\mathbf{v}}\}$ drawn from their statistical distributions. The coherence gets lost for a temporal interval $\gg \tau_c$ or a spatial interval $\gg l_c$. Here $\tau_c \sim 1/(m_av_0^2)$ and $l_c\sim 1/(m_a v_0)$ denote coherent time and coherent length of this field~\cite{Derevianko:2016vpm}, respectively, with $v_0\sim \mathcal{O}(100)\,\textrm{km/s}$ representing the halo virial velocity. For the mass range of ultralight ALDM, {\it i.e.}, $m_a \lesssim 10^{-18}\,$eV, the coherence time $\tau_c\gtrsim 20\,$years and is longer than the observation time span $T_{\rm obs}$. This implies $\omega t\approx m_a t$ in Eq.~(\ref{eq:axionf}) during the observation, and the ALDM temporal profile is thus well-described by coherent evolution of the form $\cos(m_a t+...)$.

\begin{figure} 
    \centering
        \includegraphics[width=0.9\linewidth]{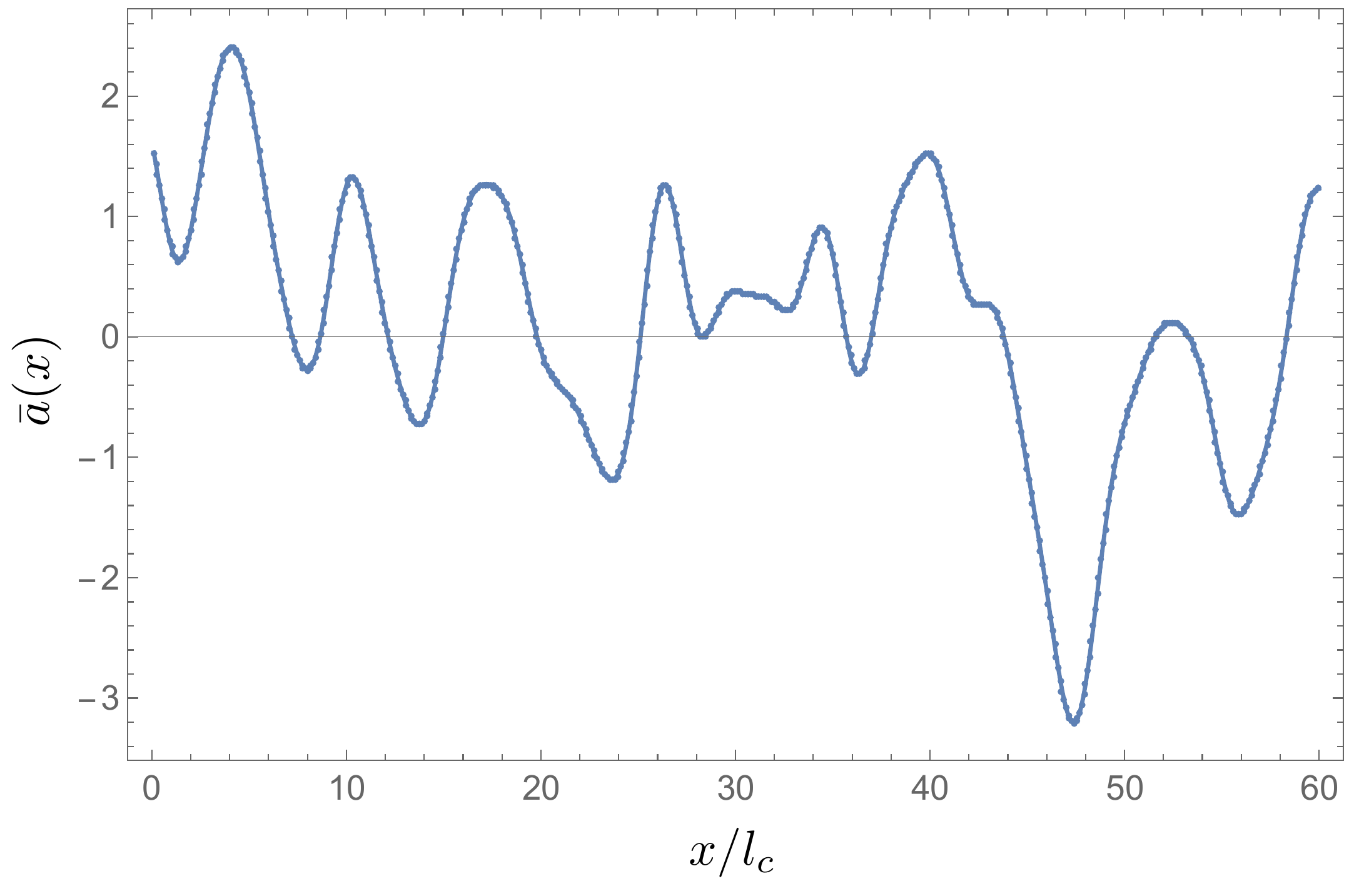}
    \caption{Spatial dependence of the normalized ALDM field $\bar{a}(x)=a(\mathbf{x},t=0)m_a/\sqrt{\rho_0}$ in the direction of $\mathbf{x}$ ($x=|\mathbf{x}|$),
    for a specific realization of the random parameters  $\{\alpha_{\mathbf{v}},\phi_{\mathbf{v}}\}$.}
    \label{fig:ALDMax}
\end{figure}

However, the pulsar distances to another pulsar or to the Earth can exceed the coherence length $l_c$. To preserve spatial information of the ALDM waves, we retain the leading term of the $\mathbf{k}\cdot \mathbf{x}$ non-relativistic expansion, namely $m_a \mathbf{v}\cdot \mathbf{x}$, in Eq.~(\ref{eq:axionf}). This term accounts for the spatial coherence within the range of coherence length as well as its loss over distances larger than $l_c$. For illustration, we show in  Fig.~\ref{fig:ALDMax} the spatial variation of the normalized ALDM field, {\it i.e.}, $\bar{a}(\mathbf{x},t)=a(\mathbf{x},t) m_a/\sqrt{\rho_0}$, for one random realization of $\{\alpha_{\mathbf{v}},\phi_{\mathbf{v}}\}$. Given the isotropy of the velocity distribution $f(\mathbf{v})$, we plot this variation in the direction of $\mathbf{x}$ ($x=|\mathbf{x}|$). The field variation approximately follows a coherent trigonometric function when $x\ll l_c$, and becomes stochastic for $x\gg l_c$. The ALDM with $m_a \sim 10^{-22}\,$eV has a coherence length of $l_c \sim \mathcal O(100)\,$pc. This number scales with $l_c\propto 1/m_a$ for other mass values. Note that this figure has been normalized and is applicable to the entire mass of ultralight ALDM.

\subsection*{Decomposition of covariant matrix}
\label{sec:pulsar_distance}

One technical challenge in evaluating the likelihood is to accurately compute the inverse of the covariance matrix. Given its high dimension and complicated structure, we employ a method to simplify this task and enhance the stability of the results.

Let us consider the covariance matrix for the red noise first, and decompose it as a matrix multiplication: 
\begin{eqnarray}
\left(\bC_{p}^r\right)_{n,m}  = ( S_{p}^r )^2(F^{r}_{p,n})^T \Phi^r_p F_{p,m}^r\,.
\end{eqnarray}
Here $\Phi^r_p$ is a $2k_{{\rm max},p}\times 2k_{{\rm max},p}$ matrix and $F^{r}_{p,n}$ is a $2k_{{\rm max},p} \times 1$ vector, and they are defined as
\begin{widetext}
\begin{equation}
\begin{aligned}
    &
    \Phi^r_p=\frac{1 }{f_{\rm yr}T_p}{\rm diag}\left\{ \left( \frac{ 1}{f_{\rm yr}T_p} \right)^{\Gamma_p}\boldsymbol{I}_{2\times 2},\dots,\left( \frac{ k_{{\rm max},p}}{f_{\rm yr}T_p} \right)^{\Gamma_p}\boldsymbol{I}_{2\times 2} \right\}\,,\qquad
    F_{p,n}^r = \begin{pmatrix}
        \cos(2\pi t_{p,n} /T_p)\\
        \sin(2\pi t_{p,n} /T_p)\\
        \vdots\\
        \cos(2\pi k_{{\rm max},p} t_{p,n} /T_p)\\
        \sin(2\pi k_{{\rm max},p} t_{p,n} /T_p)        
    \end{pmatrix}\,.
\end{aligned}
\end{equation}
\end{widetext}
$\bPhi^r_{p}$ contains the information on the spectrum shape, while $\bF^r_{p,n}$ defines the frequency patterns at the moment of $t_{p,n}$. By aggregating all observations, we define a $2k_{{\rm max},p}\times N_p$ matrix 
$F^r_p=(F^r_{p,1}, ..., F^r_{p,N_p})$, where $N_p$ is the number of observation epochs for the $p$-th pulsar. Then we can express the covariance matrix $C^r_p$ as 
\begin{eqnarray}\label{eq:Crnew}
C^r_p=( S_{p}^r )^2(F^{r}_{p})^T \Phi^r_p F_{p}^r\, ,
\end{eqnarray}
and the full matrix for $\mathcal{N}$ pulsars is: $\bC^r = {\rm diag}\{C_1^r,\,\dots,C_{\mathcal{N}}^r\}$.

Similar to the treatment of red noise, 
we decompose the signal covariant matrix $\hat{C}^a$ as a matrix multiplication: 
\begin{align}
    \hat{C}_{p,n;q,m}^a = (\bF^a_{p,n})^T\bPhi^a_{p,q} \bF^a_{q,m}\,,
\end{align}
where $\bF^a_{p,n}$ and $\bPhi^a_{p,q}$ are defined as
\begin{align}
    &\bPhi^a_{p,q} \equiv\begin{pmatrix}
\Phi^{cc}_{p,q}&\Phi^{cs}_{p,q}\\
\Phi^{sc}_{p,q}&\Phi^{ss}_{p,q}
\end{pmatrix}\, , \quad 
\bF^a_{p,n} \equiv \begin{pmatrix}
        \cos(m_at_{p,n}) \\
        \sin(m_at_{p,n})
    \end{pmatrix} \, ,
\end{align}
with 
\begin{align}
\Phi^{cc}_{p,q} &= \Phi^{ss}_{p,q} = 1 + {\rm sinc}(y_{pq}) \cos[m_a(D_p-D_q)] \\ &- {\rm sinc}(y_{ep}) \cos(m_a D_p) - {\rm sinc}(y_{eq})\sin(m_a D_q) \, , \nonumber \\
\Phi^{sc}_{p,q} &= -\Phi^{cs}_{p,q} = {\rm sinc}(y_{pq})\sin[m_a(D_p-D_q)] \\ &- {\rm sinc}(y_{ep})\sin(m_a D_p) + {\rm sinc}(y_{eq})\sin(m_a D_q) \, . \nonumber
\end{align}
Similarly, $\bF^a_{p,n}$ defines the frequency patterns at $t_{p,n}$ and $\bPhi^a_{p,q}$ contains the information on the ALDM-induced spatial correlation. We can further define a $2\times N_p$ matrix $F^a_p=(F^a_{p,1}, ..., F^a_{p,N_p})$ for individual pulsars. Then we have 
\begin{align}\label{eq:Canew}
    \hat{\bC}^a = (\bF^a)^T \bPhi^a \bF^a \, ,
\end{align}
where
\begin{eqnarray}
    \bPhi^a &=&
    \begin{pmatrix}
        \bPhi^a_{1,1}&\bPhi^a_{1,2}&\dots&\bPhi^a_{1,\mathcal{N}}\\
        \bPhi^a_{2,1}&\bPhi^a_{2,2}&\dots&\bPhi^a_{2,\mathcal{N}}\\
        \vdots & \vdots & \ddots & \vdots\\
        \bPhi^a_{\mathcal{N},1}&\bPhi^a_{\mathcal{N},2}&\dots&\bPhi^a_{\mathcal{N},\mathcal{N}}
    \end{pmatrix}\, ,\nonumber\\
    \bF^a &=& \begin{pmatrix}
        \bF^a_1&&&\\
        &\bF^a_2&&\\
        &&\ddots&\\
        &&&\bF^a_\mathcal{N} 
    \end{pmatrix} \, .
\end{eqnarray}

Finally, with the reformulated covariance matrices for both red noise and ALDM signal, 
we can apply the Woodbury matrix identity to compute the inverse of the total covariance matrix
\begin{eqnarray}    
    \bC^{-1} &=& (\bC^w + \bF^T\bPhi\bF)^{-1} \nonumber\\
    &=& (C^w)^{-1} - (C^w)^{-1} \bF^T(\bPhi^{-1}+\bF(C^w)^{-1}\bF^T)^{-1} \nonumber\\
    &&\bF(C^w)^{-1}\,   , \label{eq:covmat_inv}
\end{eqnarray}
where $\bPhi$ and $\bF$ are defined as
\begin{eqnarray}
    \bPhi &=& {\rm diag}\{(S_a)^2\Phi^a ,\, (S^r_{1})^2\Phi^r_1,\dots,\,(S^r_{\mathcal{N}})^2\Phi^r_{\mathcal{N}}\} \, \nonumber\\ 
    F &=& \begin{pmatrix}
        F^a \\
        {\rm diag}\{  F^r_{1},\,\dots,\,F^r_{\mathcal{N}}\}
    \end{pmatrix} \, .
 \label{eq:phi_all_components}
\end{eqnarray}
The determinant of the total covariance matrix is then given by
\begin{align}
    |\bC| = |\bC^w + \bF^T\bPhi\bF|  = |\bPhi^{-1}+\bF(C^w)^{-1} \bF^T| \cdot |\bPhi| \cdot |C^w| \, .
\end{align}
This method converts the evaluation of the inverse from for the high-dimensional total covariance matrix to for a matrix, namely $\bPhi$, with a much smaller dimension $2(\mathcal{N}+\sum_{p} k_{{\rm max},p})$. It thus can significantly decreases computational cost and improve numerical accuracy.

\subsection*{Statistical method}

To detect the ALDM signals, we conduct a Bayesian analysis based on the marginalized likelihood in Eq.~(\ref{Eq:Mlikelihood}):
\begin{align}
    \mathcal{L}_m = \mathcal{L}_m \left( \boldsymbol{\vartheta}|\{ \delta\PA_{p,n} \} \right)\, .
\end{align}
Here $\boldsymbol{\vartheta} = \left\{{\rm EF}_p,{\rm EQ}_p, S^r_p, \Gamma_p, D_p^*; m_a,S_a\right\}$
denotes a set of parameters, including $2\mathcal{N}$ parameters of white noise, 8 for red noise, $\mathcal{N}$ parameters of pulsar distance, and 2 for the ALDM signal. Then, for a given ALDM mass $m_a$, we derive the 95\% upper limits on $S_a$ which are further converted to the ones on the ALDM Chern-Simons coupling $g_{a\gamma\gamma}$.

\begin{figure*}[ht]
    \centering
    \includegraphics[scale=0.7]{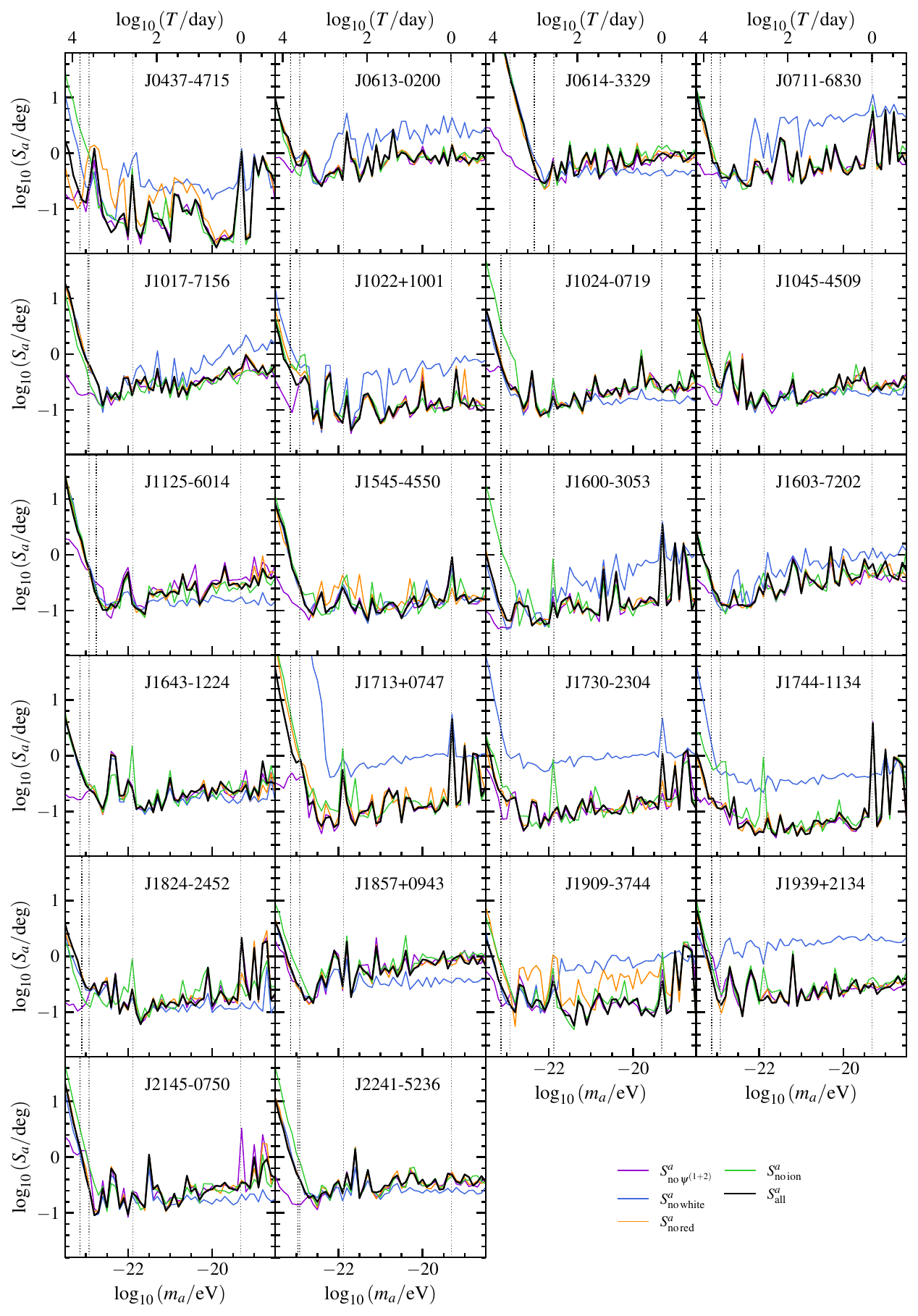}
    \caption{95\% upper limits on the ALDM characteristic signal strength $S_a$ as a function of $m_a$ for each of the 22 PPPA MSPs. The notation follows that of Fig.~\ref{fig:Salimit}. 
    }
    \label{fig:Salimit_supp}
\end{figure*}

Fig.~\ref{fig:Salimit_supp} displays the 95\% upper limits on $S_a$ as a function of $m_a$  for each of the 22 PPPA MSPs, supplementing Fig.~\ref{fig:Salimit} in the main text. The difference between the black and colored solid lines reflects the impacts of individual noise components on the $S_a$ limits. For each component, such impacts may vary a lot among different pulsars. Specifically, the ionospheric subtraction using \texttt{ionFR} effectively removes sharp peaks around $\sim 1 \,$yr for J1600--3053, J1643--1224, J1730--2304, J1744--1134, J1824--2452 and J1939+2134, and reduces excess low-frequency noises for J0437--4715, J1024--0719, J1600--3053 and J1744--1134. Incorporating the additional contributions of white noise described by the parameters EF$_p$ and EQ$_p$ in Eq.~(\ref{eq:new_sigma}) strengthens the limits significantly from J1713+0747, J1730--2304, J1744--1134, J1939+2134, etc. Incorporating the red noise component notably improves the limits from J0437--4715 and J1909--3744 at both intermediate and low frequencies. 
To account for long-term deterministic effects, we conservatively apply a second-order polynomial model to all pulsars, although strong evidence of linear and quadratic temporal variations has been observed for some of them only. This strategy results in less stringent limits at frequencies below $1/T_{\rm obs}$, a small portion of the frequency range of interest. 
Overall, for the most sensitive pulsars, such as J0437--4715, J1744--1134, etc., these outcomes imply that using the minimal noise model may result in an underestimation of sensitivity by several times and even over an order of magnitude.

The PPA response to a true ALDM signal can be manifested as Bayes factors with 
the full-correlation signal model against the null-signal model (${\rm BF}^{\rm Full}_{\rm Null} \equiv Z^{\rm Full}/Z^{\rm Null}$) and the full-correlation signal model against the auto-correlation-only signal model (${\rm BF}^{\rm Full}_{\rm Auto} \equiv Z^{\rm Full}/Z^{\rm Auto}$). Here $Z^{\rm Full}$, $Z^{\rm Null}$ and $Z^{\rm Auto}$ are statistical evidence for these three models. 
For the calculation of these Bayes factors with the MCMC chains, we use the product-space method ~\cite{carlin1995bayesian,lodewyckx2011tutorial} when the values are small, and the thermodynamical integral~\cite{calderhead2009estimating} when the values are large.

To demonstrate this point, we perform a mock study by applying the PPA analysis to the noise-only model and the noise plus ALDM signal models, respectively. The former case  includes only white and red noises, with the model parameter values derived from fitting  real data. The latter case incorporates the injected signals additionally. Here, we consider two ALDM benchmark scenarios: $\{m_a, g_{a \gamma\gamma}\}$ = $\{10^{-21} \, {\rm eV}, 10^{-12} {\rm GeV}^{-1} \}$ and $\{10^{-23} \, {\rm eV}, 10^{-14} {\rm GeV}^{-1} \}$, with the corresponding characteristic signal strength $S_a$ being approximately $0.1\,$deg for both.  

\begin{figure*}[h]
    \centering
    \includegraphics[width=1\linewidth]{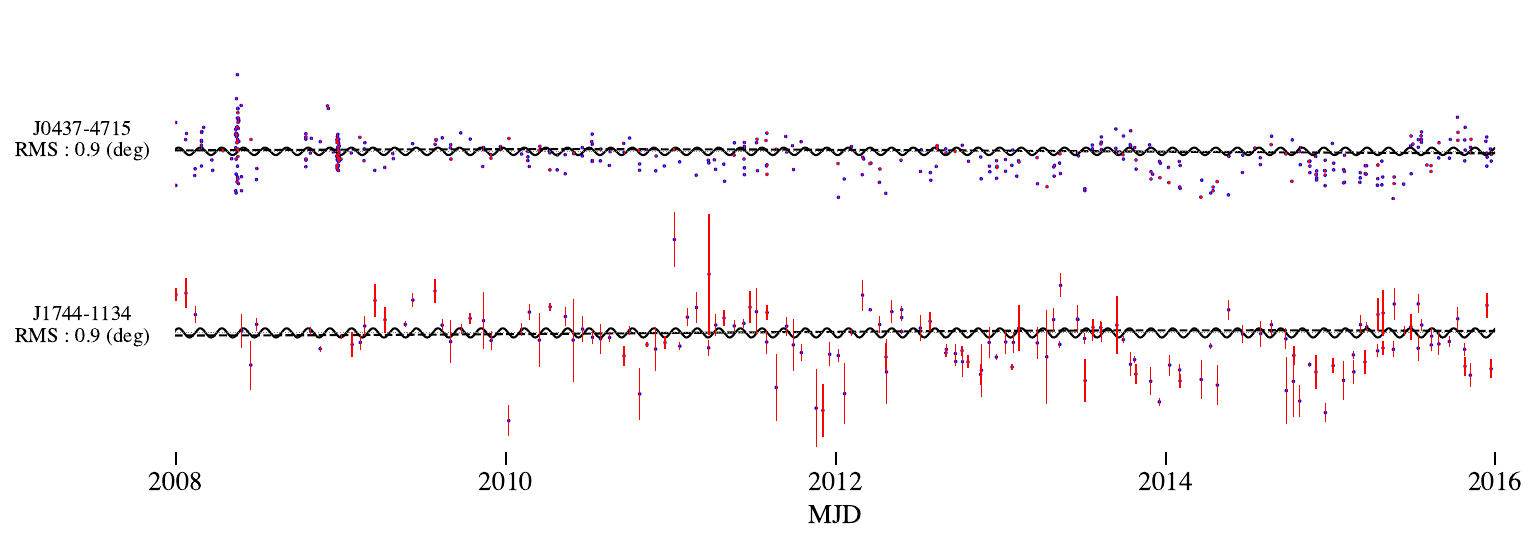}
    \caption{ALDM-induced $\Delta$PA time series  alongside the observed PA residual time series from Fig.~\ref{fig:dPA} for the two most sensitive PPPA pulsars, namely J0437--4715 and J1744--1134. The solid and dashed wavy lines represent the ALDM benchmark scenarios with $\{m_a, g_{a \gamma\gamma}\}$ = $\{10^{-21} \, {\rm eV}, 10^{-12} {\rm GeV}^{-1} \}$ and $\{10^{-23} \, {\rm eV}, 10^{-14} {\rm GeV}^{-1} \}$, respectively, where $S_a\approx 0.1\,$deg for both. }
    \label{fig:PADataSignal}
\end{figure*}

We present the ALDM-induced $\Delta$PA time series for these two scenarios in Fig.~\ref{fig:PADataSignal}, alongside the observed PA residual time series from Fig.~\ref{fig:dPA} for the two most sensitive PPPA pulsars, J0437--4715 and J1744--1134. These signals oscillate with different characteristic frequencies. But their amplitude or strength is smaller than the RMS of the observed PA residuals for these pulsars, making it challenging to probe these ALDM scenarios. However, as demonstrated below using mock data, the PPA Bayesian analysis, together with the aforementioned Bayes factors, can successfully detect and recognize these signals by correlating the 22 pulsars. The relevant noise and signal parameters can be reasonably recovered also.

\begin{figure*}[ht]
    \centering
        \includegraphics[width=0.35\linewidth]{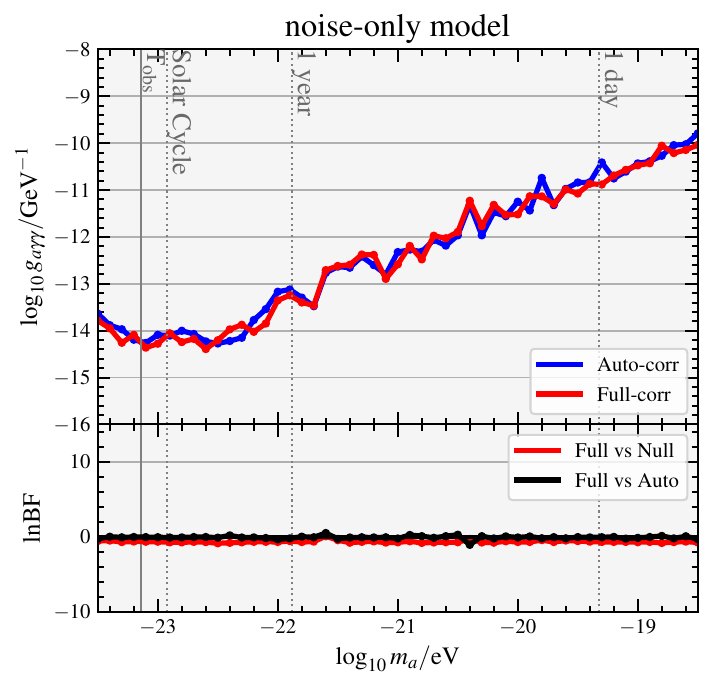} \\
            \includegraphics[width=0.35\linewidth]{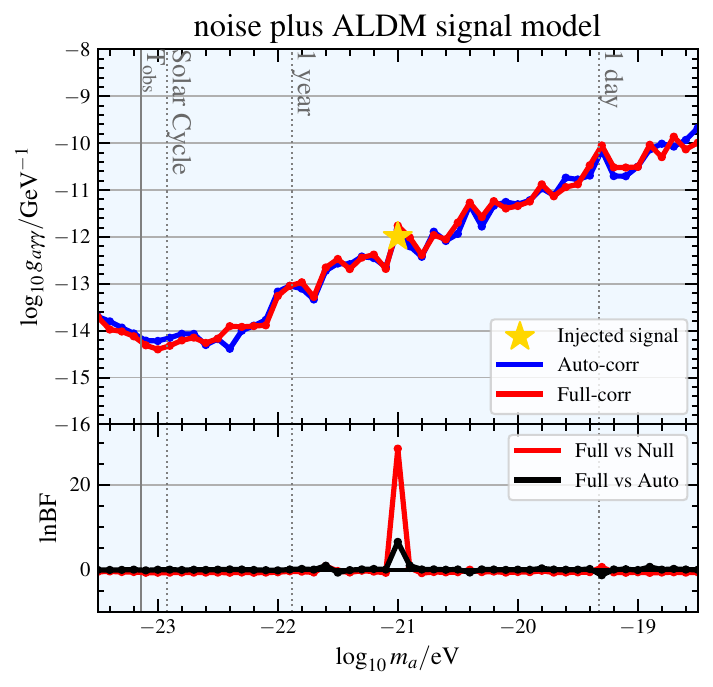}
         \includegraphics[width=0.35\linewidth]{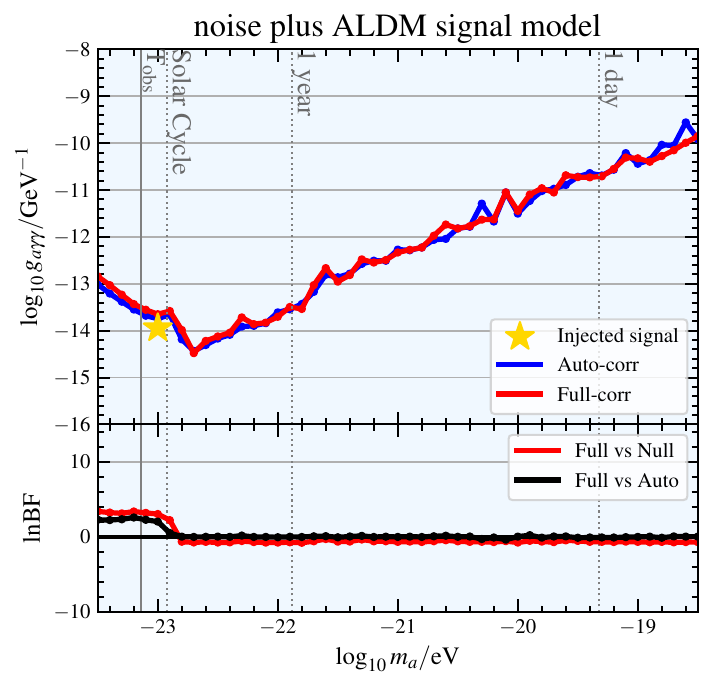}
    \caption{
     Mock PPA limits on the ALDM Chern-Simons coupling $g_{a\gamma\gamma}$ and mock performance of the Bayes factors $\ln{\rm BF}^{\rm Full}_{\rm Null}$ and $\ln{\rm BF}^{\rm Full}_{\rm Auto}$, as a function of the ALDM mass $m_a$. 
     The top panel displays a sample of random noise only, 
     while the bottom panels show two samples with the injected ALDM signals from the benchmark scenarios defined in Fig.~\ref{fig:PADataSignal}. Here the yellow stars indicate the corresponding $\{m_a, g_{a\gamma\gamma}\}$ values for these two scenarios.
    The vertical solid line denotes the PPPA observation period $T_{\rm obs}\approx 18~{\rm yrs}$, and the dotted lines for other reference timescales, including solar cycle, one year and one day.}
    \label{fig:mock_1}
\end{figure*}

\begin{figure*}[t]
    \centering
    \includegraphics[width=0.75\linewidth]{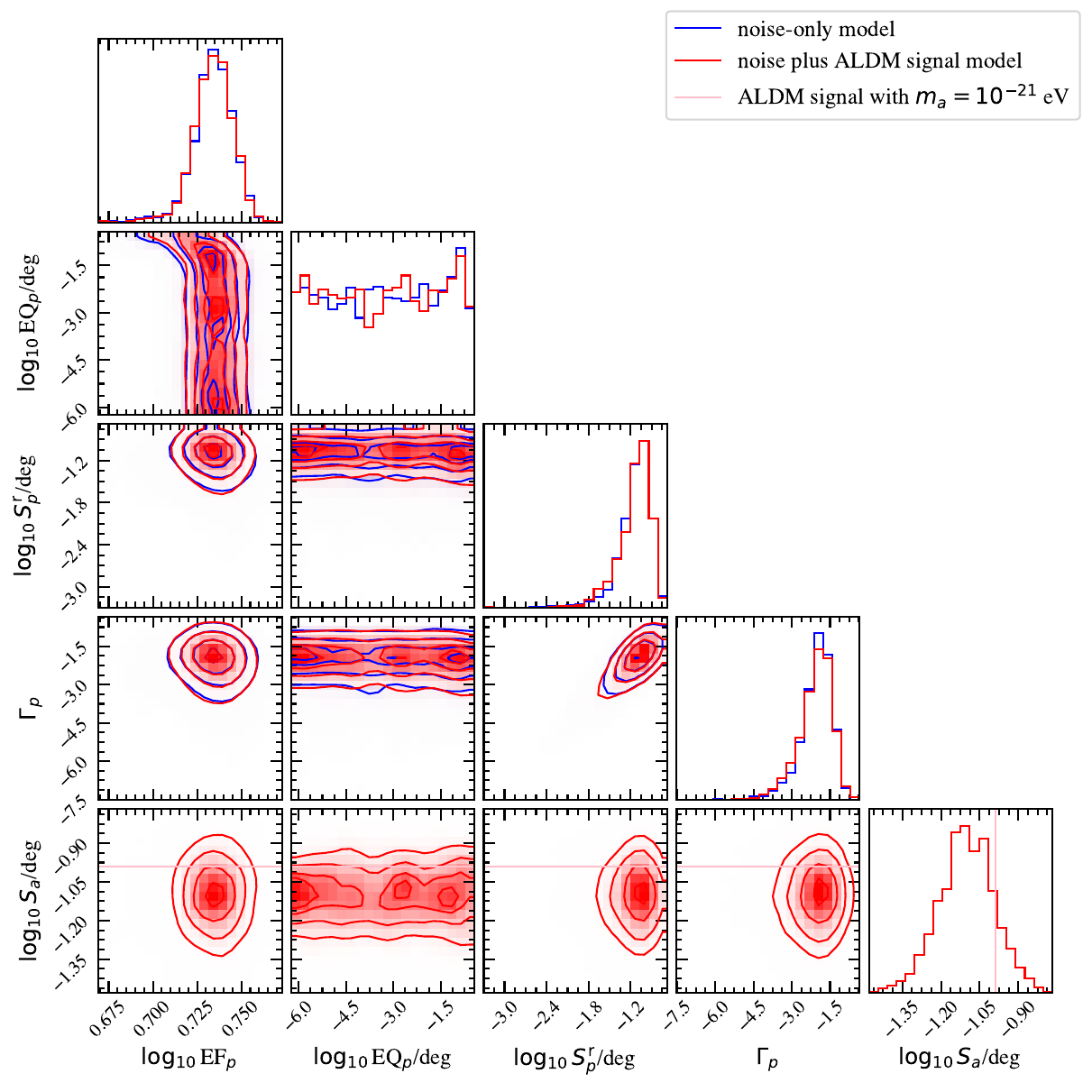}
    \caption{Mock posterior distributions for the random noise parameters of the most sensitive pulsar, {\it i.e.}, J0437--4715, and the characteristic signal strength $S_a$ from the PPA analysis at $m_a=10^{-21}\,$eV. The red contours, corresponding to the bottom-left panel of Fig.~\ref{fig:mock_1}, represent the ALDM benchmark scenario with $\{m_a, g_{a \gamma\gamma}\}$ = $\{10^{-21} \, {\rm eV}, 10^{-12} {\rm GeV}^{-1} \}$ (where $S_a\approx 0.1\,$deg), while the blue contours illustrate the results for the same sample but with no ALDM signals injected. The vertical or horizontal pink lines indicate the $S_a$ value for the injected ALDM signals.
    }
    \label{fig:posterior}
\end{figure*}

We first illustrate in Fig.~\ref{fig:mock_1} the PPA limits on $g_{a\gamma\gamma}$ and the $\ln{\rm BF}^{\rm Full}_{\rm Null}$ and $\ln{\rm BF}^{\rm Full}_{\rm Auto}$ responses to the ALDM signals for the considered samples of mock data.
For the mock data containing the random noises only (top panel), both Bayes factors fluctuate around zero, indicating no clear evidence for a signal with either full correlation or auto-correlation only. However, with the ALDM signals injected, these two Bayes factors demonstrate some dramatical deviation from the baseline of $\ln {\rm BF}\sim 0$, exceeding the detection threshold of 2.3 for strong evidence.
If the ALDM mass is larger than $1/T_{\rm obs}$, allowing the signal to be well-resolved, a sharp peak appears in both Bayes factors at the true mass value (bottom-left panel). 
Building on this, the strength of injected signals $S_a$ indeed can be  recovered with a high precision, as illustrated by the mock posterior distributions in Fig.~\ref{fig:posterior}, which exhibit minimal degeneracy with noise parameters.
If the ALDM mass is comparable to or smaller than $1/T_{\rm obs}$, the signal cannot be fully resolved, resulting in an upward deformation in both Bayes factors below the true mass value (bottom-right panel). As 
$\ln{\rm BF}^{\rm Full}_{\rm Null}$ measures the deviation from the null signal scenario and $\ln{\rm BF}^{\rm Full}_{\rm Auto}$ examines the role of cross correlation in recognizing the signal, these features can serve as key signatures for the ALDM detection.
These discussions and observations clearly indicate that the ALDM signals can be efficiently detected through the PPA analysis, despite the incorporation of  additional components into the full noise model. We emphasize that this outcome is to be expected, given the unique correlation pattern of the ALDM signals which can differentiate them from the noise.

\begin{figure*}[ht]
    \centering
    \includegraphics[scale=0.65]{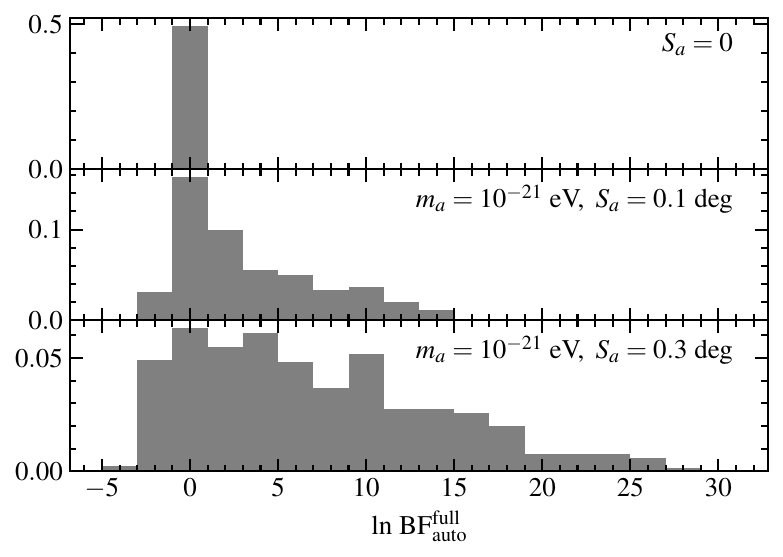}
    \caption{Mock distributions of the Bayes factor $\ln{\rm BF}^{\rm Full}_{\rm Auto}$ in three different signal scenarios. In the top panel, the mock data are generated with white and red noises only, based on parameter values derived from fitting the real data. In the middle and bottom panels, a stochastic ALDM signal is injected into these random noises, with $m_a=10^{-21}\,$eV and $S_a = 0.1\,$deg and $0.3\,$deg, respectively. 600 random realizations are analyzed for each case,  all at $m_a=10^{-21}\,$eV.}
    \label{fig:mock_constraint}
\end{figure*}

So far, only one specific realization for random  noises and stochastic ALDM signals has been applied for the analysis of mock data. To statistically support this argument, we particularly display in Fig.~\ref{fig:mock_constraint} the mock distributions of the Bayes factor $\ln{\rm BF}^{\rm Full}_{\rm Auto}$ for three different signal scenarios, based on totally 1800 random realizations of both noises and signals. For the case with random noises only, the Bayes factor distribution, which is calculated at $m_a = 10^{-21}\,$eV for the convenience of discussions, is centered around zero.
In the cases with injected signals, we consider an  ALDM mass larger than $1/T_{\rm obs}$, namely $m_a = 10^{-21}\,$eV, for demonstration. The Bayes factor $\ln{\rm BF}^{\rm Full}_{\rm Auto}$ is then derived at this mass value to plot the distributions. As the signal strength, {\it i.e.}, the $S_a$ value,  increases, the distribution shifts to the positive side, with increased median and variance. This outcome further highlights the role of $\log{\rm BF}^{\rm Full}_{\rm Auto}$ or the pulsar cross-correlation in recognizing the real signals in the data.

We present the PPPA limits on the ALDM Chern-Simons coupling in Fig.~\ref{fig:results} for the full-correlation and auto-correlation-only signal models. Both sets of limits display sharp peaks on the timescales of one day or shorter and at $m_a \approx 10^{-22}\,$eV. To acquire insights regarding the nature of these peaks, we demonstrate both Bayes factors, namely $\ln{\rm BF}^{\rm Full}_{\rm Null}$ and $\ln{\rm BF}^{\rm Full}_{\rm Auto}$, as functions of $m_a$ in the same figure. Notably, while the peaks also appear in the $\ln{\rm BF}^{\rm Full}_{\rm Null}$ curve on the similar timescales, $\ln{\rm BF}^{\rm Full}_{\rm Auto}$ remains near zero across the entire mass range (similar to the $\ln{\rm BF}^{\rm Full}_{\rm Auto}$ shown in the upper panel of Fig.~\ref{fig:mock_1} and also consistent with its distribution shown in the upper panel of Fig.~\ref{fig:mock_constraint}). 
This suggests that no evidence for the ALDM-induced cross-correlation or signal has been found in the analyzed data, and the sharp peaks observed in the limits and also the $\log{\rm BF}^{\rm Full}_{\rm Null}$ curve are more likely caused by imperfect noise modeling for individual pulsars.


\bibliography{ref.bib}

\begin{thebibliography}{103}
\bibitem[{foo()}]{footnote1}
\bibinfo{note}{One pulsar, J1824-2452, is located closer to the galactic center
  than the others in the PPPA, where the DM density is potentially higher.
  However, given the limited sensitivity of this pulsar to the ALDM signals, we
  do not anticipate a significant impact on the combined limits which are
  derived under the assumption of a uniform $\rho(\mathbf{x})$ for all PPPA
  pulsars.}
\end{thebibliography}

\newoutputstream{stream}
\openoutputfile{counters5}{stream}
\addtostream{stream}{
  \protect\setcounter{equation}{\arabic{equation}}}
\addtostream{stream}{
  \protect\setcounter{table}{\arabic{table}}}  
\addtostream{stream}{
  \protect\setcounter{figure}{\arabic{figure}}}    
\closeoutputstream{stream}

\end{document}